\documentclass{aa} 

\usepackage{graphicx}
\usepackage{txfonts}
\usepackage{tikz}
\usepackage[colorlinks=true,urlcolor=blue,citecolor=blue,
            linkcolor=blue]{hyperref}
\usepackage{mhchem}

\def\aap{A\&A}

\def\apjl{ApJL}
\def\apjs{ApJS}

\def\la{\mathrel{\mathchoice {\vcenter{\offinterlineskip\halign{\hfil
$\displaystyle##$\hfil\cr<\cr\sim\cr}}}
{\vcenter{\offinterlineskip\halign{\hfil$\textstyle##$\hfil\cr
<\cr\sim\cr}}}
{\vcenter{\offinterlineskip\halign{\hfil$\scriptstyle##$\hfil\cr
<\cr\sim\cr}}}
{\vcenter{\offinterlineskip\halign{\hfil$\scriptscriptstyle##$\hfil\cr
<\cr\sim\cr}}}}}
\def\ga{\mathrel{\mathchoice {\vcenter{\offinterlineskip\halign{\hfil
$\displaystyle##$\hfil\cr>\cr\sim\cr}}}
{\vcenter{\offinterlineskip\halign{\hfil$\textstyle##$\hfil\cr
>\cr\sim\cr}}}
{\vcenter{\offinterlineskip\halign{\hfil$\scriptstyle##$\hfil\cr
>\cr\sim\cr}}}
{\vcenter{\offinterlineskip\halign{\hfil$\scriptscriptstyle##$\hfil\cr
>\cr\sim\cr}}}}}

\def\pat{p_{\rm atom}}
\def\pgas{p_{\rm gas}}
\def\pH2O{{p_{\rm H2O}}}
\def\pCO2{{p_{\rm CO2}}}
\def\pCH4{{p_{\rm CH4}}}
\def\pN2{{p_{\rm N2}}}
\def\pNH3{{p_{\rm NH3}}}
\def\pO2{{p_{\rm O2}}}
\def\pHH{{p_{\rm H2}}}

\def\epsC{\epsilon_{\rm C}}
\def\epsO{\epsilon_{\rm O}}

\def\bff{\sffamily\bfseries}

\begin{document} 

%\title{Characterisation of exoplanet atmospheres based on their H, C
%  and O element abundances}
\title{Coexistence of CH$_4$, CO$_2$ and H$_2$O in exoplanet atmospheres}

\author{P.~Woitke\inst{1,2}
       \and
        O.~Herbort\inst{1,2,3}
       \and
        Ch.~Helling\inst{1,2,4}
       \and
        E.~St{\"u}eken\inst{1,3}
       \and
        M.~Dominik\inst{1,2}
       \and
        P.~Barth\inst{1,2,3}
       \and
        D.~Samra\inst{1,2}
}

\institute{
           Centre for Exoplanet Science, University of St Andrews, 
           St Andrews, UK
         \and
           SUPA, School of Physics \& Astronomy, University of St
           Andrews, St Andrews,  KY16 9SS, UK 
         \and
           School of Earth \& Environmental Studies, University of St
           Andrews, St Andrews, KY16 9AL, UK
         \and
           SRON Netherlands Institute for Space Research, Sorbonnelaan 2,
           3584 CA Utrecht, NL
}

\date{Received 08/07/2020; accepted 17/10/2020}

\abstract{We propose a classification of exoplanet atmospheres based
  on their H, C, O, N element abundances {below about} 600\,K.
  Chemical equilibrium models were run for all combinations of H, C,
  N, O abundances, and three types of solutions were found, which are
  robust against variations of temperature, pressure and nitrogen
  abundance. Type~A atmospheres contain \ce{H2O}, \ce{CH4}, \ce{NH3}
  and either \ce{H2} or \ce{N2}, but only traces of \ce{CO2} and
  \ce{O2}. Type~B atmospheres contain \ce{O2}, \ce{H2O}, \ce{CO2} and
  \ce{N2}, but only traces of \ce{CH4}, \ce{NH3} and \ce{H2}.  Type~C
  atmospheres contain \ce{H2O}, \ce{CO2}, \ce{CH4} and \ce{N2}, but
  only traces of \ce{NH3}, \ce{H2} and \ce{O2}.  Other molecules are
  only present in ppb or ppm concentrations in chemical equilibrium,
  depending on temperature.  Type~C atmospheres are not found in the
  solar system, where atmospheres are generally cold enough for water
  to condense, but exoplanets may well host such atmospheres. Our
  models show that graphite (soot) clouds can occur in type~C
  atmospheres {in addition to water clouds}, which can occur in all
  types of atmospheres.  Full equilibrium condensation models show
  that the outgassing from warm rock can naturally provide type~C
  atmospheres.  We conclude that type~C atmospheres, if they exist,
  would lead to false positive detections of biosignatures in
  exoplanets when considering the coexistence of \ce{CH4} and
  \ce{CO2}, and suggest other, more robust non-equilibrium markers.}

\keywords{planets and satellites: atmospheres --
          planets and satellites: composition --
          astrochemistry 
         }
         % planets and satellites: physical evolution --

\maketitle

%---------------------------------------------------------------------------
\section{Introduction}
%---------------------------------------------------------------------------

The detection of exoplanets that exhibit spectral signatures of
biological activity (biosignatures) is one of the most urgent goals of
modern astronomy.  Because of observational limitations, such
biosignatures are currently based on certain combinations of abundant
molecules that exhibit detectable spectroscopic signatures at medium
spectral resolution in the infrared, such as \ce{H2O}, \ce{CO2},
\ce{CH4} and \ce{CO}. Other potentially abundant molecules like
\ce{H2}, \ce{O2} and \ce{N2} have no permanent dipole moment and are
more difficult to detect. Molecules generally need to have a minimum
concentration of about $10^{-4}$ (100\,ppm) to be detectable with JWST
\citep{KrissansenTotton2019,Sousa-Silva2020}, and here a number of
candidates have been discussed recently.  For example, \ce{O3} as an
indicator for the presence of \ce{O2} \citep{Gaudi2018},
\ce{SO2}-derived sulphate aerosols as an indicator for volcanic
activity \citep{Misra2015}, or \ce{PH3} as biosignature
\citep{Sousa-Silva2020}.

Searching for suitable combinations of detectable molecules that
suggest biological activity, \citet{KrissansenTotton2019} have
recently proposed to consider the coexistence of \ce{CO2} and \ce{CH4}
(without CO) in planetary atmospheres, as is true for Earth
\citep{Meadows2018}. In fact, CO$_2$ and CH$_4$ represent the two
endpoints of the redox-spectrum of carbon, and it is hence not obvious
why both molecules should be present simultaneously.
\citet{Krissansen-Totton2018} argued that a disequilibrium between
\ce{CH4} and \ce{CO2}, accompanied by \ce{N2} and liquid \ce{H2O}, was
present during the Archean on Earth. \citet{Sandora2020} review the
evolutionary processes involving \ce{CH4} in the Earth atmosphere.
Once the specific geological processes cease that used to cause that
disequilibrium, it seems unlikely that non-biological processes could
maintain such an atmosphere on habitable exoplanets.  However,
Krissansen-Totton et al.\ did not explore in how far \ce{CO2} and \ce{CH4} 
can simply coexist in chemical equilibrium.

Previous chemical models in astronomy have mainly focused on H-rich
atmospheres. For example, \citet{Moses2013a} and \citet{Hu2014} have
presented sparse grids of chemical models varying the H abundance,
metallicity and C/O-ratio to study the atmospheric composition of mini
Neptune and super Earths, including kinetical quenching,
photodissociation and vertical mixing.  These papers show that carbon
can mostly form \ce{CO2} at low H abundances, rather than \ce{CO} and
\ce{CH4} as used from the \ce{H2}-dominated atmospheres.
\citet{Heng2016} have considered an analytic nine-molecule model in
chemical equilibrium to discuss the effects of varying C/O and N/O
ratios on \ce{H2}-dominated hot exoplanet atmospheres, aiming at a
fast tool to retrieve the atmospheric composition from exoplanet
observations.  \citet{Morley2017a} have considered chemical
equilibrium models for the atmospheres of Earth-sized exoplanets
including those of the TRAPPIST-1 system, based on element abundances
taken from Venus, Earth and Titan, to discuss their observability with
JWST.

Simulating the chemical evolution of the Earth atmosphere,
\citep{Zerkle2012} have identified the photo-dissociation of \ce{CH4}
as important non-equilibrium physical process in the upper
atmosphere. The low dissociation energy of \ce{CH4}
($\sim\!\!4.3$\,eV) implies that this molecule can be dissociated
quite easily compared to \ce{CO2}. Therefore, UV irradiation is
thought to significantly change the atmospheric \ce{CH4}/\ce{CO2}
ratio over time \citep{Zerkle2012}, and if the hydrogen atoms (or
\ce{H2} molecules) can escape, which are produced by this reaction,
the photo-dissociation of \ce{CH4} is able to remove hydrogen from
exoplanet atmospheres, creating more oxidising conditions.
\citet{Arney2015} proposed that during the Archean, Earth was likely
covered in photochemical haze triggered by this process. Another large
uncertainty here is the Earth's history of surface pressure and
atmospheric \ce{N2} abundance, which was possibly affected by the
evolution of life as well \citep{Stuecken2016}.

The aim of this paper is to present an exhaustive study of the
composition of exoplanet atmospheres considering all possible
combinations of H, C, N and O abundances assuming chemical
equilibrium.  Substantial deviations from chemical equilibrium can be
caused by biological activity, but other physical and geological
processes may explain these disequilibria, too, for example UV and
cosmic ray irradiation.  But, in any case, potential biosignatures
should not be based on combinations of molecules which are already
expected in chemical equilibrium \citep{Seager2013}.

%\citet{Cockell2019}: habitability as a series of binary questions,
%questioned by \citet{Heller2020}.

\begin{figure*}
  \vspace*{-2mm}
  \resizebox{182mm}{!}{
  \begin{minipage}{185mm}
  \begin{tabular}{ccc}
  \hspace*{-5mm}
  \includegraphics[height=59mm,trim=10 0 79 11,clip]{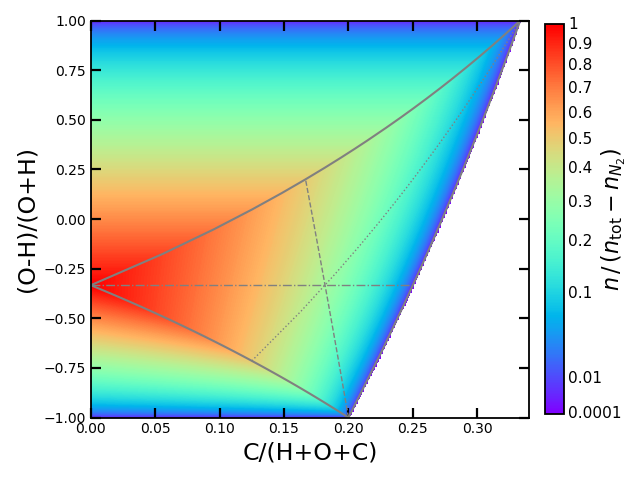} &
  \hspace*{-5mm}
  \includegraphics[height=59mm,trim=64 0 79 11,clip]{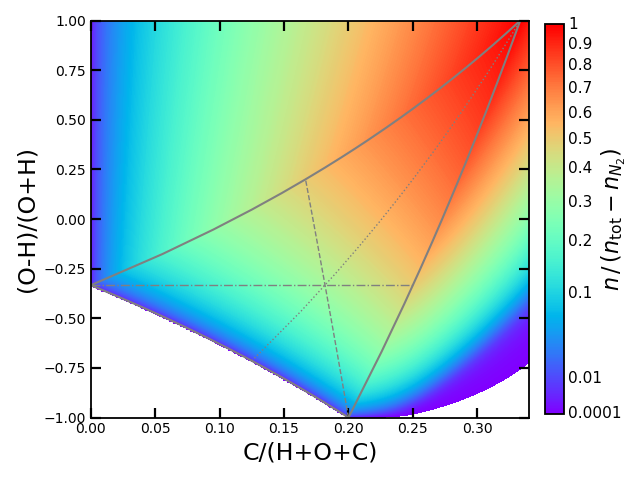} &
  \hspace*{-5mm}
  \includegraphics[height=59mm,trim=64 0  0 11,clip]{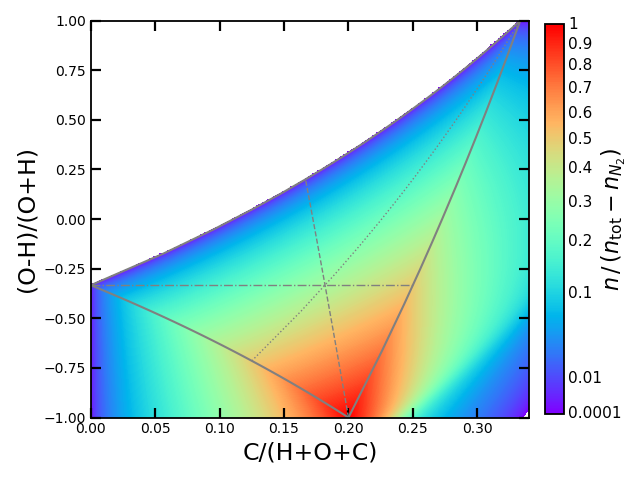} \\[-6.5mm]
  \hspace*{-5mm}
  \includegraphics[height=59mm,trim=10 0 79 11,clip]{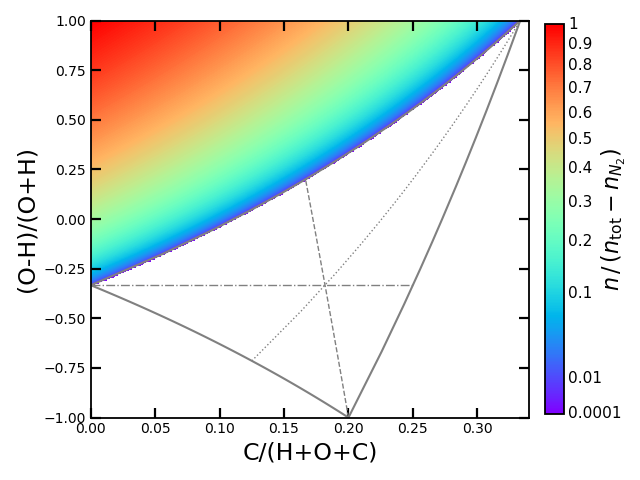} &
  \hspace*{-5mm}
  \includegraphics[height=59mm,trim=64 0 79 11,clip]{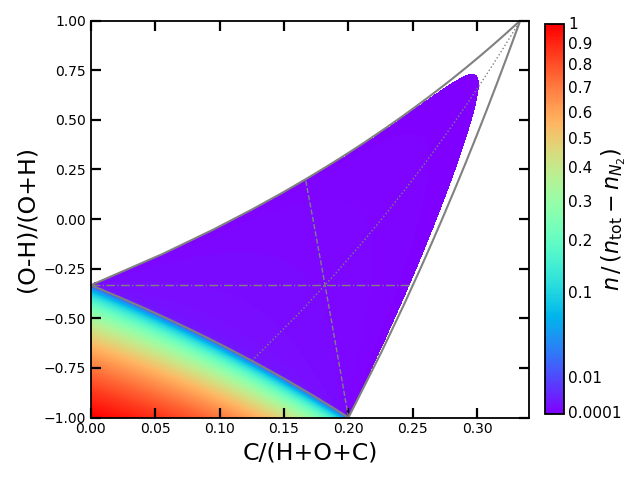} &
  \hspace*{-5mm}
  \includegraphics[height=59mm,trim=64 0  0 11,clip]{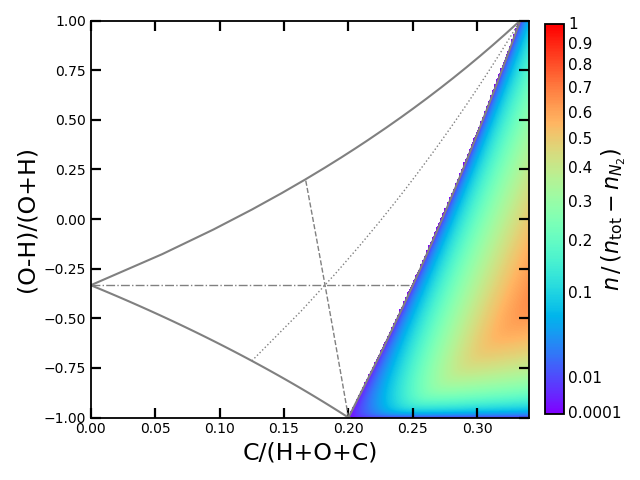} 
  \end{tabular}\\[-85mm]
  \hspace*{10mm}{\large\bff\color{white}H$_{\mathbf 2}$O}\\[-30mm]
  \hspace*{105mm}{\large\bff\color{white}CO$_{\mathbf 2}$}\\[37mm]
  \hspace*{148mm}{\large\bff\color{white}CH$_{\mathbf 4}$}\\[9mm]
  \hspace*{10mm}{\large\bff\color{white}O$_{\mathbf 2}$}\\[39mm]
  \hspace*{66mm}{\large\bff\color{white}H$_{\mathbf 2}$}\\[-13mm]
  \hspace*{168mm}{\large\bff\color{black}CO}\\[13mm]
  \end{minipage}}
  \caption{Molecular concentrations in chemical equilibrium as
    function of H, C and O element abundances, calculated for
    $T\!=\!400$\,K and $p\!=\!1$\,bar. The central grey triangle marks
    the region within which \ce{H2O}, \ce{CH4} and \ce{CO2} coexist in
    chemical equilibrium.  The thin grey lines indicate where two
    concentrations are equal: $n_{\ce{H2O}}\!=\!n_{\ce{CO2}}$ (dashed),
    $n_{\ce{CO2}}\!=\!n_{\ce{CH4}}$ (dash-dotted), and $n_{\ce{CH4}}\!=\!n_{\ce{H2O}}$
    (dotted). $n$ means particle densities.  Blank regions have
    concentrations $<\!10^{-4}$. In order to enhance the colour
    contrasts, a colour-map was chosen that is linear in
    $(n/(n_{\rm tot}-n_{\ce{N_2}}))^{1/2}$ from 0.01 to 1.}
  \label{fig:6mol}
  \vspace*{-2mm}
\end{figure*}

%---------------------------------------------------------------------------
\section{Simplified chemical equilibrium}
\label{ChemEquil}
%---------------------------------------------------------------------------
All known atmospheres of solar system bodies are mainly composed of
hydrogen (H), oxygen (O), carbon (C) and nitrogen (N), so we will
focus on the H\,--\,C\,--\,N\,--\,O system in this paper, which is the most
pressing system for the observational characterisation of exoplanets
and the identification of possible biosignatures.
 
\begin{table}[!b]
\vspace*{-3mm}
\centering
\caption{Atomisation energies $E_a$ of selected molecules}
\label{tab:Ea}
\vspace*{-2mm}
\resizebox{7.7cm}{!}{\begin{tabular}{c|c||c|c}
\hline
molecule & $E_a$ at $T\!=\!0\,K$ & molecule & $E_a$ at $T\!=\!0\,K$\\
& \\[-2.2ex]
\hline
& \\[-2.2ex]
 \ce{CH4} & 17.02\;eV & \ce{ CO} & 11.11\;eV \\
 \ce{C2H2}& 16.86\;eV & \ce{ N2} &  9.76\;eV \\
 \ce{CO2} & 16.56\;eV & \ce{H2O} &  9.51\;eV \\
 \ce{HCN} & 13.09\;eV & \ce{ O2} &  5.12\;eV \\
 \ce{NH3} & 12.00\;eV & \ce{ H2} &  4.48\;eV \\ 
%\ce{ NO} &  6.51\;eV & \ce{ CN} &  7.77\;eV \\
\hline
\end{tabular}}
\vspace*{-1mm}
\end{table}

At low temperatures, $T\!\la\!600\,$K, the Gibbs free energies
$\Delta G_f = \Delta H_f - T \Delta S$ are dominated by the enthalpy
of formation $\Delta H_f$, which means that in chemical equilibrium,
the molecular concentrations will essentially minimise $\Delta H_f$.
In Table~\ref{tab:Ea} we list a few atomisation energies, i.e.\ the
energies required to convert molecules into neutral atoms, for
example $E_a(\ce{H2O}) = 2\Delta H_f(\ce{H})+\Delta H_f(\ce{O})-\Delta
H_f(\ce{H2O})$. The data have been extracted from the NIST-Janaf
tables \citep{Chase1982,Chase1986}.

Simple combinatorics shows that \ce{CH4}, \ce{CO2},
\ce{H2O} and \ce{N2} are the thermodynamically most favourable
molecules to minimise $\Delta H_f$. For example, all of
the following reactions are exothermic\\*[-2mm]
%%% ($\Delta H_f<0$),\\*[-2mm]
\begin{equation*}
  \begin{array}{rcll}
  \ce{H2}   + \frac{1}{4}\;\ce{CO2} &\longrightarrow& 
    \frac{1}{2}\;\ce{H2O} + \frac{1}{4}\;\ce{CH4}
    &:\rm -0.39\,eV\\[0.8mm]
  \ce{CO}   + \frac{1}{2}\;\ce{H2O} &\longrightarrow& 
    \frac{3}{4}\;\ce{CO2} + \frac{1}{4}\;\ce{CH4}
    &:\rm -0.81\,eV\\[0.8mm]
  \ce{HCN}  + \frac{3}{4}\;\ce{H2O} &\longrightarrow& 
    \frac{3}{8}\;\ce{CO2} + \frac{5}{8}\;\ce{CH4} + \frac{1}{2}\;\ce{N2} 
    &:\rm -1.51\,eV\\[0.8mm]
  \ce{C2H2} + \frac{3}{2}\;\ce{H2O} &\longrightarrow& 
    \frac{3}{4}\;\ce{CO2} + \frac{5}{4}\;\ce{CH4}              
    &:\rm -2.56\,eV\\[0.8mm]
  \ce{NH3}  + \frac{3}{8}\;\ce{CO2} &\longrightarrow& 
    \frac{3}{4}\;\ce{H2O} + \frac{3}{8}\;\ce{CH4} + \frac{1}{2}\;\ce{N2} 
    &:\rm -2.83\,eV\\[0.8mm]
%  \ce{NO}   + \frac{1}{4}\;\ce{CH4} &\longrightarrow& 
%    \frac{1}{2}\;\ce{H2O} + \frac{1}{4}\;\ce{CO2} + \frac{1}{2}\;\ce{N2}    
%    &:\rm -3.01\,eV\\[0.8mm]
  \ce{O2}   + \frac{1}{2}\;\ce{CH4} &\longrightarrow& 
             \;\;\;\ce{H2O} + \frac{1}{2}\;\ce{CO2} 
    &:\rm -4.17\,eV
%  \ce{CN}   +        \;\;\;\ce{H2O} &\longrightarrow& 
%    \frac{1}{2}\;\ce{CO2} + \frac{1}{2}\;\ce{CH4} + \frac{1}{2}\;\ce{N2}
%    &:\rm -4.34\,eV \ ,
  \end{array}
\end{equation*}                        
which means that all other molecules can react exothermally to
eventually form a mixture of only \ce{CH4}, \ce{CO2}, \ce{H2O} and
\ce{N2} to minimise the Gibbs free energy at low temperatures. The
first reaction requires to break a C-O bond, likely biologically
mediated, and has been named ``methanogenesis'' by \citet{Woese1977}
and \citet{Waite2017}. With increasing temperature,
entropy becomes more relevant, and \ce{H2} is the first among the
trace molecules to reach significant concentrations, followed by CO.

Therefore, we first consider a planetary atmosphere in which the most
abundant molecules are \ce{H2O}, \ce{CH4}, \ce{CO2} and \ce{N2}
{(henceforth called type~C atmospheres)}, such
that the total gas pressure is approximately given by
\begin{equation}
  \pgas ~\approx~ \pH2O + \pCH4 + \pCO2 + \pN2 \ .
  \label{pgas}
\end{equation}
The element conservation equations can be expressed in terms of a
fictitious total pressure after complete atomisation, $\pat$,
here $\pat=3\,\pH2O+5\,\pCH4+3\,\pCO2+2\,\pN2$,
\begin{eqnarray}
  H\cdot\pat ~=~ 2\,\pH2O + 4\,\pCH4 \label{totH} &\!\!\!,&
  C\cdot\pat ~=~ \pCO2 + \pCH4\ ,\\[-0.5mm]
  O\cdot\pat ~=~ \pH2O + 2\,\pCO2\hspace*{2mm} &\!\!\!,&
  N\cdot\pat ~=~ 2\,\pN2 \ ,
  \label{totC}
\end{eqnarray}
where $H$, $C$, $O$ and $N$ are the given element abundances
normalised by the condition $H+C+O+N=1$. The usual element abundances
in astronomy are defined with respect to hydrogen, for example
$\epsC=C/H$ and $\epsO=O/H$, but
here we wish to also consider hydrogen-poor atmospheres with
$H\!\to\!0$.
%For example, $C\,\pat$ is the total number density of carbon nuclei in
%the gas phase multiplied by $kT$.

\begin{figure*}
  \vspace*{-1mm}
  \begin{tabular}{cc}
  $N\!=\!0.001$ & $N\!=\!0.5$\\[-1mm]
  \hspace*{-3mm}
  \includegraphics[width=92mm,trim=10 12 5 11,clip]{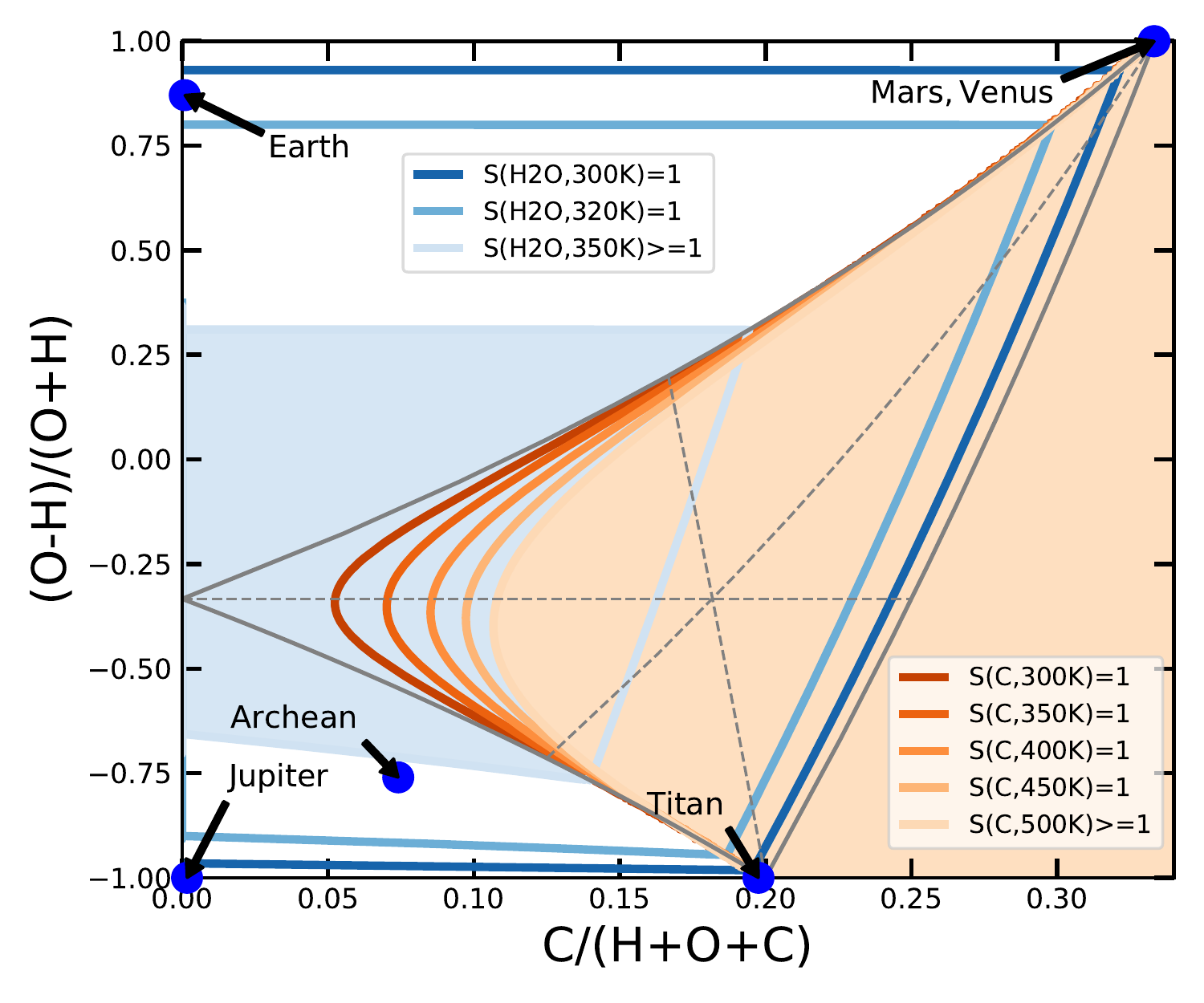}
  &
  \hspace*{-5mm}
  \includegraphics[width=92mm,trim=10 12 5 11,clip]{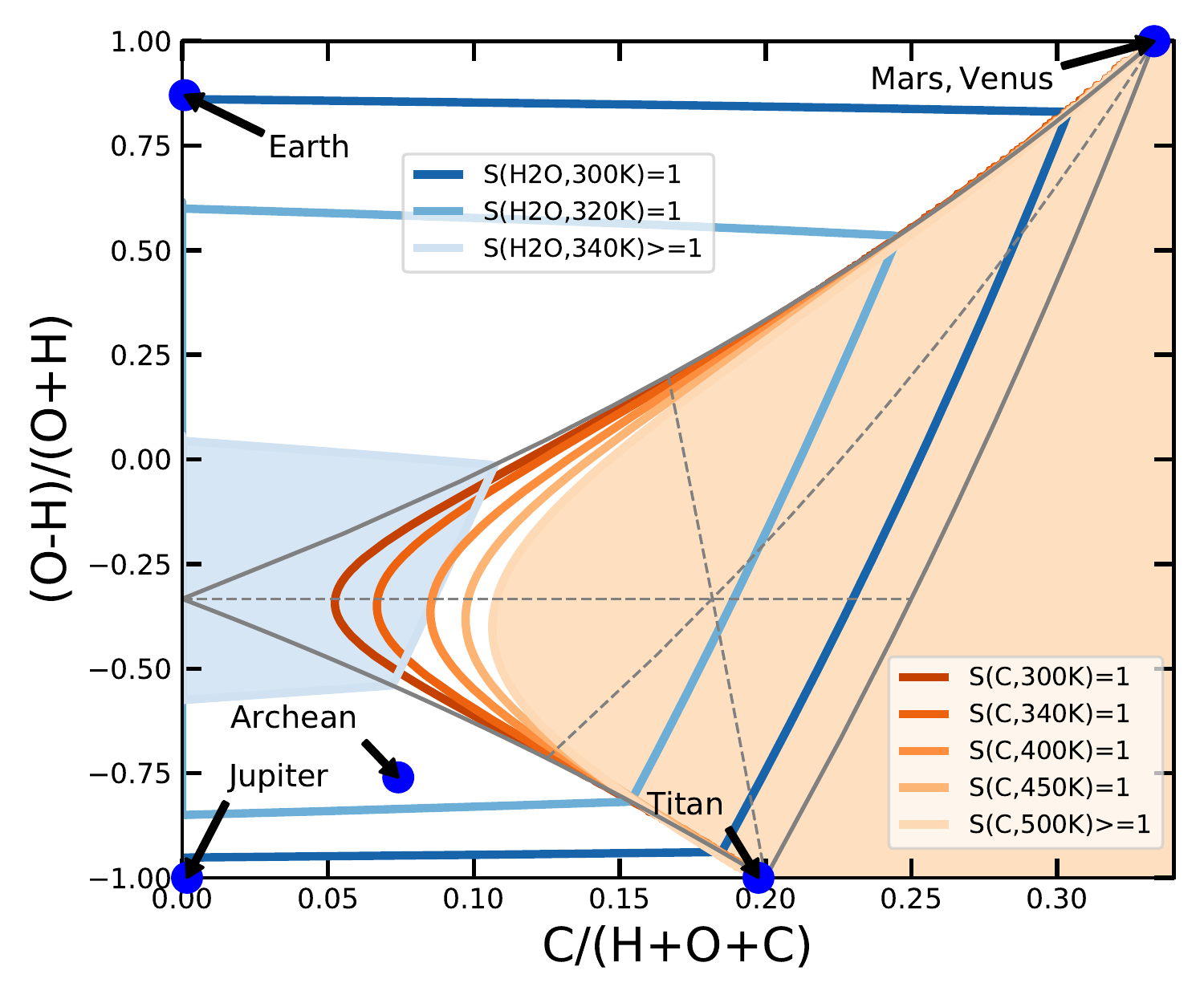}
  \end{tabular}
  \vspace*{-2mm}
  \caption{Impact of liquid water (\ce{H2O}) and graphite (C)
    condensation at constant pressure of $1$\,bar.  The blue and
    orange contour lines mark where the supersaturation ratio $S$
    equals one for water and graphite, respectively, at selected
    temperatures. Inside of the light blue and light orange shaded
    regions, the gas is supersaturated with respect to water at 350\,K
    and graphite at 500\,K, respectively. For higher temperatures,
    lower pressures or higher $N$ abundances, the shaded areas shrink
    and eventually vanish.  The left diagram is calculated for a small
    nitrogen abundance $N\!=\!0.001$, whereas $N\!=\!0.5$ is assumed
    in the right diagram. The atmospheric compositions of Earth, Mars,
    Venus, Jupiter, Titan and the Archean Earth \citep[taken
      from][]{Miller1953} are marked. Earth (assuming 1.5\% water
    content) sits right on top of the $S(\ce{H2O},300\,{\rm
      K},N\!=\!0.5)\!=\!1$ line.}
  \label{fig:Sats}
  \vspace*{-2mm}
\end{figure*}

This system of linear equations, Eqs.\,(\ref{pgas}) to (\ref{totC}),
can be readily solved for the molecular partial pressures
\begin{equation}
  \begin{array}{ccc}
  \displaystyle \frac{\pH2O}{\pgas} = \frac{H+2\,O-4\,C}{H+2\,O+2\,N}
     \hspace*{2.5mm} &\hspace*{-1mm},\hspace*{-1mm}&
  \displaystyle \frac{\pCO2}{\pgas} =
  \frac{2\,O+4\,C-H}{2\,H+4\,O+4\,N} \ ,\\[4mm]
  \displaystyle \frac{\pCH4}{\pgas} = \frac{H-2\,O+4\,C}{2\,H+4\,O+4\,N}
                     &\hspace*{-1mm},\hspace*{-1mm}&\hspace*{-0.7mm}
  \displaystyle \frac{\pN2}{\pgas}  = \frac{2\,N}{H+2\,O+2\,N} \ .
  \label{psol}
  \end{array}
\end{equation}
%As an illustrative example, let us consider a water dominated
%exoplanet atmosphere with $H\!=\!0.6$, $O\!=\!0.3$, $C\!=\!0.1$ and $N\!=\!0$,
%in which case we find the following partial pressures
%\begin{equation*}
%  \pH2O = 0.667\,\pgas \;\;;\;\;
%  \pCO2 = 0.167\,\pgas \;\;;\;\;
%  \pCH4 = 0.167\,\pgas \ .
%\end{equation*}
%Thus, CO$_2$ and CH$_4$ can coexist in exoplanet atmospheres in chemical
%equilibrium. 
%This somewhat surprising result is a consequence of our assumption
%that only three molecules H$_2$O, CH$_4$ and CO$_2$ are relevant
%as carriers of the three elements hydrogen, carbon and oxygen,
%resulting in a linear equation system where these elements {\sl
%  must} be distributed this way, just because of the stoichiometry of
%the molecules. 
%which allow for a further analysis of the equilibrium between
%\ce{H2O}, \ce{CO2} and \ce{CH4} in planetary atmospheres.  
If any of the partial pressures according to
Eq.\,(\ref{psol}) is negative, we have obviously left the region of
applicability of our assumptions.  These side conditions are
\begin{eqnarray}
  O > 0.5\,H + 2\,C\hspace*{4.3mm}
  \label{tri-1}
  &\Rightarrow& \mbox{\ce{O2}-rich atmosphere, no \ce{CH4}}\\
  H > 2\,O + 4\,C\hspace*{7mm}  
  \label{tri-2}
  &\Rightarrow& \mbox{\ce{H2}-rich atmosphere, no \ce{CO2}}\\
  C > 0.25\,H + 0.5\,O  
  \label{tri-3}
  &\Rightarrow& \mbox{\ce{graphite\ condensation}, no \ce{H2O}}\ .
\end{eqnarray}
% If there is a surplus of oxygen after subtraction of the maximum
% possible amount of oxygen contained in \ce{H2O}, which is given by
% $H/2$, and the maximum possible amount contained in \ce{CO2}, which is
% given by $2\,C$ (Eq.~\ref{tri-1}), we get an $\ce{O2}$-rich atmosphere
% where \ce{CH4} is only present in trace amounts.  Similarly, if there
% is a surplus of hydrogen, after subtraction of the hydrogen in
% \ce{H2O} and \ce{CH4} (Eq.~\ref{tri-2}), we get a \ce{H2}-rich
% atmosphere with only traces of \ce{CO2}, and if there is a surplus of
% carbon, after subtraction of \ce{CH4} and \ce{CO2} (Eq.~\ref{tri-3}),
% there would be pure carbon in the gas phase, which is highly
% sutersaturated and is expected to form graphite.
We can furthermore use Eqs.~(\ref{psol}) to determine
where two of the molecular partial pressures are equal
\begin{eqnarray}
      H = 2\,O \hspace*{9mm}       &\quad\Longleftrightarrow\quad 
                      & \pCO2 = \pCH4 \label{equal-1}\\ 
  12\,C = 2\,O + 3\,H &\quad\Longleftrightarrow\quad 
                      & \pCO2 = \pH2O \label{equal-2}\\ 
  12\,C = 6\,O + H \hspace*{2.3mm}   &\quad\Longleftrightarrow\quad 
                      & \pH2O = \pCH4 \label{equal-3}\ .
\end{eqnarray}
The partial pressure of \ce{N2} is always positive according to
Eq.\,(\ref{psol}), providing no additional constraints. Hence,
nitrogen does not interfere with the H\,--\,C\,--\,O system, unless
there is a surplus of $H$ according to Eq.\,(\ref{tri-2}), in which
case \ce{NH3} becomes an abundant molecule, see Sect.~\ref{typeA}.

\section{Full chemical equilibrium models}

To confirm the simplified analysis presented in
Sect.\,\ref{ChemEquil}, we ran full gas-phase chemical equilibrium
models with {\sc GGchem} \citep{Woitke2018} for the elements H, C, N,
and O. {\sc GGchem} finds 52 molecular species in its database for
this element mixture: $\ce{H2}$, $\ce{C2}$, $\ce{N2}$, $\ce{O2}$,
$\ce{CH}$, $\ce{NH}$, $\ce{OH}$, $\ce{CN}$, $\ce{CO}$, $\ce{NO}$,
$\ce{HCN}$, $\ce{CHNO}$, $\ce{HCO}$, $\ce{CH2}$, $\ce{H2CO}$,
$\ce{CH3}$, $\ce{CH4}$, $\ce{CNO}$, $\ce{CNN}$, $\ce{NCN}$,
$\ce{CO2}$, $\ce{C2H}$, $\ce{C2H2}$, $\ce{C2H4}$, $\ce{C2H4O}$,
$\ce{C2N}$, $\ce{C2N2}$, $\ce{C2O}$, $\ce{C3}$, $\ce{C3O2}$,
$\ce{C4}$, $\ce{C4N2}$, $\ce{C5}$, $\ce{HNO}$, $\ce{HONO}$,
$\ce{HNO2}$, $\ce{HNO3}$, $\ce{HO2}$, $\ce{NH2}$, $\ce{N2H2}$,
$\ce{H2O}$, $\ce{NH3}$, $\ce{N2H4}$, $\ce{NO2}$, $\ce{NO3}$,
$\ce{N2O}$, $\ce{N2O3}$, $\ce{N2O4}$, $\ce{N2O5}$, $\ce{N3}$,
$\ce{O3}$ and $\ce{C3H}$. The results for $T\!=\!400$\,K and
$p\!=\!1$\,bar are shown in Fig.~\ref{fig:6mol}.  The region of
coexistence of \ce{H2O}, \ce{CO2} and \ce{CH4} is indicated by a grey
triangle corresponding to Eqs.\,(\ref{tri-1}) to (\ref{tri-3}), and
the dashed lines of equal concentration correspond to
Eqs.\,(\ref{equal-1}) to (\ref{equal-3}). The models have been
computed with a small nitrogen abundance $N\!=\!10^{-3}$, but the
results are independent of $N$ when plotted as function of $C/(H+O+C)$
and $(O-H)/(O+H)$, since nitrogen does not interfere significantly
with the H\,--\,C\,--\,O system. See Appendix~\ref{sec:types} for a
discussion in how far our results depend on temperature, pressure, and
nitrogen abundance.

These results show that, indeed, none of the other molecules are
relevant to this problem.  \ce{N2}, \ce{H2O}, \ce{CO2} and \ce{CH4}
are the only molecules that must be considered within the grey
triangle to solve the element conservation equations, at least
approximately, and therefore the argumentation presented in
Sect.\,\ref{ChemEquil} holds.

\begin{table}
\vspace{0.5mm}
\caption{Trace gas concentrations in chemical equilibrium with element
  abundances $N\!=\!2/13$, $H\!=\!6/13$, $O\!=\!3/13$ and $C\!=\!2/13$,
  where main constituents are 25\% \ce{N2}, 25\% \ce{H2O}, 25\%
  \ce{CO2} and 25\% \ce{CH4}.}
\label{contaminations}
\vspace*{-2mm}
\resizebox{90mm}{!}{\begin{tabular}{c|cccccc}
\hline
 &&&&&\\[-2ex]
 & 200\,K & 300\,K & 400\,K & 500\,K & 600\,K & 700\,K \\
 &&&&&\\[-2.3ex]
\hline
 &&&&&\\[-2ex]
 \ce{H2}  & 2\,ppb   & 5.8\,ppm & 390\,ppm & 0.52\%  & 2.9\%    & 9.5\% \\[1mm]
 \ce{CO}  & $<$\,1\,ppb& $<$\,1\,ppb& 260\,ppb & 39\,ppm & 0.11\%   & 1.2\% \\[1mm]
 \ce{NH3} & 205\,ppb & 4.7\,ppm & 23\,ppm  & 58\,ppm & 101\,ppm & 130\,ppm \\[1mm]
\hline
\end{tabular}}
\vspace*{-3mm}
\end{table}

The patterns shown in Fig.~\ref{fig:6mol} are robust against changes
of pressure, temperature, and nitrogen abundance, see
Appendix~\ref{sec:types} for details.  This is the true strength of
this diagram. All results can be approximately inferred just from the
identification and stoichiometry of the four most stable molecules,
making these results suitable for a classification of exoplanet
atmospheres.  The central triangle shows trace concentrations of
\ce{H2} on a level of a few $10^{-4}$ at 400\,K, otherwise the gas
composition is very pure in chemical equilibrium, with only three
abundant molecules at any point besides \ce{N2}. With increasing
temperature, the \ce{H2} trace concentration increases, followed by
the occurrence of \ce{CO} in trace concentrations, see
Table~\ref{contaminations}.  Molecules not listed in
Table~\ref{contaminations} have even lower abundances. We conclude
that the simplified analysis of the coexistence of \ce{H2O}, \ce{CO2}
and \ce{CH4} as presented in Sect.\,\ref{ChemEquil} is valid to about
600\,K.

The results of \citet{Moses2013a} and \citet[][see their
  Figs.\,4-6]{Hu2014} show similar patterns, but their sparse grid of
models and usage of $H$-abundance and C/O ratio make it difficult to
compare their results with ours in detail.

%---------------------------------------------------------------------------
\section{Water and graphite condensation}
%---------------------------------------------------------------------------
{\sc GGchem} also computes the supersaturation ratios of graphite
\ce{C[s]}, liquid water \ce{H2O[l]}, solid water \ce{H2O[s]}, and the
ices of ammonia \ce{NH3[s]}, methane \ce{CH4[s]}, \ce{CO[s]} and
\ce{CO2[s]}.  While the ices only deposit at $T\la\!200$\,K
at 1\,bar, the effect of carbon and water condensation is
significant, see Fig.~\ref{fig:Sats}. Shaded areas in this diagram
indicate that at least one condensed species is supersaturated
$(S\!>\!1)$. Such gases are expected to form clouds, and the
precipitation of cloud particles would remove the respective
elements from the gas phase.  Exoplanet atmospheres can hence not
reside within the shaded areas, but are expected to move toward the
edges of the blank regions in Fig.~\ref{fig:Sats}, where they come to
rest.

However, the extent of the supersaturated areas depends on
temperature, pressure and nitrogen abundance, and hence on atmospheric
height, which complicates the analysis.  Earth, for example, is
not a point in Fig.~\ref{fig:Sats}, but a line, because the
water content in the gas phase increases with temperature, guided by
the condition $S(\ce{H2O},T)\!\approx\!1$.

Fig.~\ref{fig:Sats} shows that at $350\,$K, 1\,bar and small nitrogen
abundance, the triangle in which \ce{H2O}, \ce{CO2} and \ce{CH4}
coexist, is mostly covered by the supersaturated areas, either
graphite or water or both.  At $400\,$K, however, water cannot
condense anywhere in the left diagram, and then the grey triangle of
coexistence is only partly inhibited by graphite condensation. The
grey triangle is less affected by supersaturation for lower pressures
or increased nitrogen abundances. The occurrence of graphite clouds in
exoplanet atmospheres was discussed, for example, by \citet[][see
  their Fig.~7]{Moses2013a}.

\begin{figure}
  \hspace*{-0.5mm}
  \includegraphics[width=90mm,height=80mm,trim=0 3 0 28,clip]{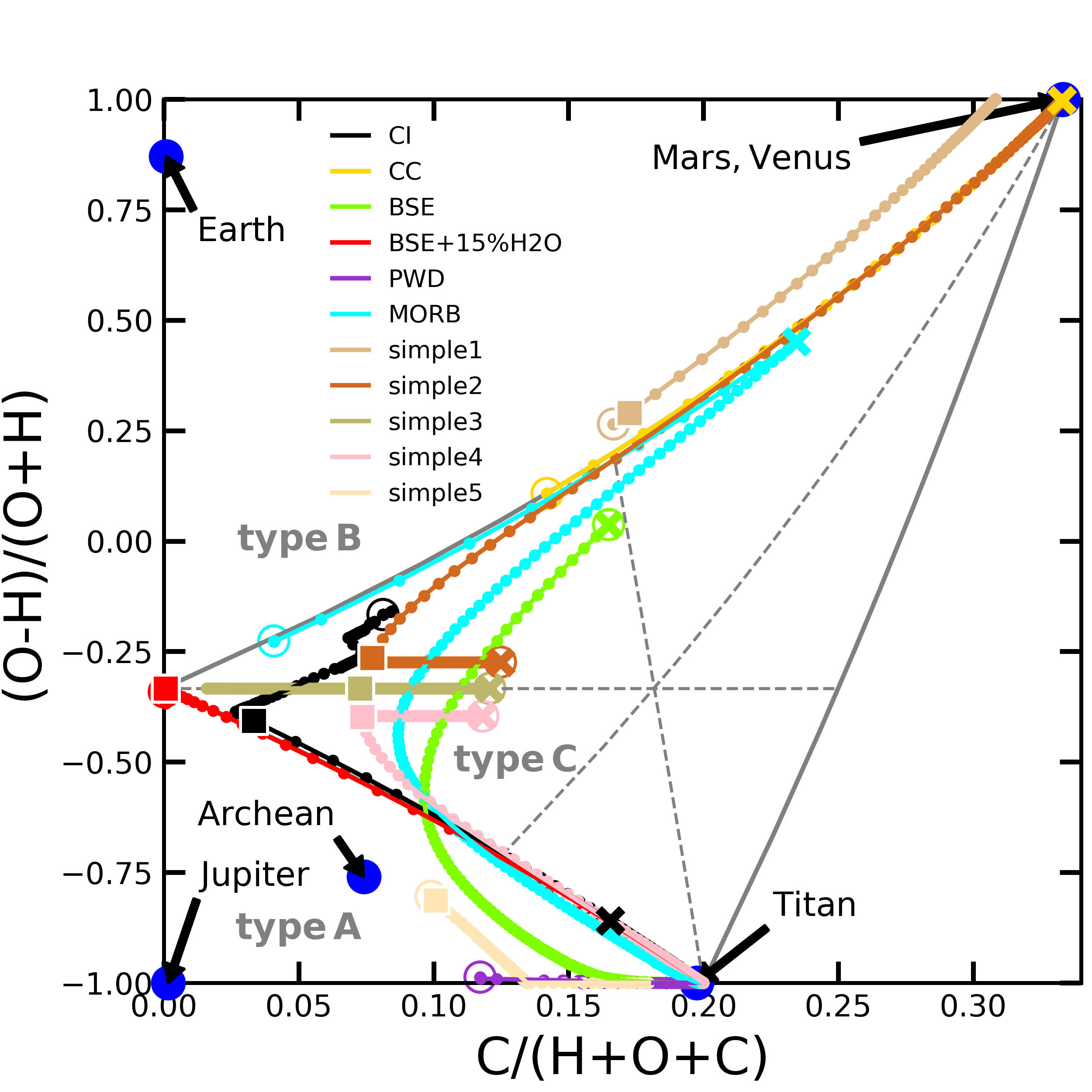}
  \vspace*{-5mm}
  \caption{%Results of equilibrium condensation models at 1\,bar.
    %The basis coloured contours show the logarithmic concentration of
    %\ce{CO2} which is more or less independent of temperature
    %(computed for $T\!=\!400\,$K). Blank regions indicate regions
    %where the concentration is $<\!10^{-4}$.  
    Results from six full equilibrium condensation models for 18
    elements from \citet{Herbort2020}, re-computed for $p\!=\!1\,$bar,
    and five simple models for just 4 elements, see
    Table~\ref{SimpleModels}.  Each model starts at $T\!=\!600\,$K
    (marked by circles) and then we follow that model with a line
    down to 200\,K with 200 log-equidistant points. Abbreviations are
    carbonaceous chondrites (CI), Mid Oceanic Ridge Basalt (MORB),
    Continental Crust (CC), Bulk Silicate Earth (BSE), and abundances
    deduced from Polluted White Dwarf observations (PWD). The points
    where the trajectories start to have graphite are marked by
    crosses, and the points where liquid water starts to occur are
    marked by squares.}
    %(only CI and BSE+15\%\ce{H2O})
    %``no\_phl'': phyllosilicates are not included in the model. }
  \label{fig:full}
  \vspace*{-1mm}
\end{figure}

Figure \ref{fig:full} shows the results of some full equilibrium
condensation models for 18 elements from \citet{Herbort2020}.  In
these models, the total ($=$ condensed $+$ gas phase) element
abundances are taken from different materials found in the Earth
crust, meteorites and polluted white dwarfs, see explanation of
abbreviations in the figure caption. The model determines which liquid
and solid materials are present at each temperature and subtracts the
respective condensed element fractions, until $S\!\leq\!1$ is achieved
for all condensates. The remaining gas phase element abundances are
plotted in Fig.~\ref{fig:full} with lines, where we start at 600\,K
(circle) and follow the results down to 200\,K. The changes of the H,
O and C abundances along these tracks are caused by the effects of
progressive condensation in these models, in particular
phyllosilicates, carbonates, graphite and water.  The CC model is very
dry and eventually creates an almost pure \ce{N2} atmosphere with some
\ce{CO2} at low temperatures.  All other models eventually form a
mixture of \ce{N2} and \ce{CH4}, similar to Titan's atmosphere. In the
MORB and BSE models, graphite condenses below 550\,K and 600\,K,
respectively, but there is no liquid water. The PWD model shows
neither graphite nor water condensation, but the CI and the
water-enriched BSE models show liquid water at 369\,K and 373\,K,
respectively, below which the models follow tracks, due to the removal
of \ce{H2O} and C, along the borderline between type A and C, where the
\ce{CO2}-concentration is low but still notable, for example the CI
model at 350\,K has 10\% \ce{N2}, 48\% \ce{CH4}, 42\% \ce{H2O} and
0.35\% \ce{CO2}.

In Fig.\,\ref{fig:full}, we have additionally overplotted some simple
equilibrium condensation models for the four element H, C, N, O only, where
the initial element abundances are arbitrarily set as listed in
Table~\ref{SimpleModels}.  These models show the principle behaviour
of type A, B and C atmospheres when only affected by water and
graphite condensation. At $\sim\!350$\,K and 1\,bar, the three type~C models
show similar abundances of \ce{CO2} and \ce{CH4}, both with
concentrations of a few percent, besides the major molecules \ce{N2}
and \ce{H2O}, next to liquid water.

\section{Conclusions}

A mixture of \ce{H2O}, \ce{CO2}, \ce{CH4} and \ce{N2} has been found
to be the most favourable combination of molecules to minimise the
Gibbs free energy at low temperatures $T\!\la\!600\,$K.  If all
available elements can be converted into these molecules, the gas will
only contain these species, whereas all other molecules only have
trace concentrations in chemical equilibrium.  However, if this is
not possible, due to stoichiometric constraints, additional types of
atmospheres occur:
\begin{enumerate}
\item[A)] Type~A atmospheres are $H$-rich and mostly contain
  \ce{CH4}, \ce{H2O} and \ce{NH3}, but \ce{CO2} and \ce{O2} are
  lacking.  In Appendix~\ref{typeA} we show that the forth abundant
  molecule is either \ce{H2} (type~A1) or \ce{N2} (type~A2). The
  atmospheres of Jupiter, Titan and Archean Earth belong to type~A.
  Titan's cold atmosphere is almost water-free because of
  condensation, but does have some \ce{H2} (about 0.15\%,
  \citealt{Niemann2005}).
\item[B)] Type~B atmospheres are $O$-rich and mostly contain
  \ce{O2}, \ce{N2}, \ce{CO2} and \ce{H2O}, but \ce{CH4}, \ce{NH3} and
  \ce{H2} are lacking.  Earth belongs to this type. The Martian
  atmosphere is a members of type~B, too, because it contains some
  \ce{O2} (about 0.1\%\footnote{See fact-sheets at
    \url{https://nssdc.gsfc.nasa.gov/planetary}.\label{foot1}}), Venus
  is too hot for our classification, but its atmosphere is also mostly
  made of \ce{CO2} and \ce{N2}, with traces of \ce{H2O}, but no
  \ce{CH4} and no \ce{NH3}$^{\ref{foot1}}$.
\item[C)] Type~C atmospheres mostly contain \ce{H2O}, \ce{CO2}, \ce{CH4} and \ce{N2},
  but \ce{NH3}, \ce{H2} and \ce{O2} are lacking.  This type
  of atmospheres was discussed in Sect.\,\ref{ChemEquil}.
\end{enumerate}
The molecule \ce{CO} is not abundant in any type of atmosphere at low
temperatures in chemical equilibrium.

Type~C exoplanet atmospheres 
% could also be named water-atmospheres, because \ce{H2O} can be the
% most abundant species in these atmospheres.  Such atmospheres
are not found in the solar system, possibly because the low
temperatures of the objects favour water condensation, which reduces the hydrogen and
oxygen abundances, however, slightly warmer exoplanets may well host
such atmospheres. Type~C atmospheres are featured by the coexistence
of \ce{CH4} and \ce{CO2}, both with percent-concentrations, which is
possible in equilibrium only in type~C atmospheres. Our models
furthermore suggest that only in type~C atmospheres, carbon can
directly condense to form graphite (soot) clouds.

\begin{table}
\vspace*{1mm}
\caption{Setup of simple equilibrium condensation models for four
  elements. $T_{\rm C}$ and $T_{\rm H2O}$ are the temperatures below
  which graphite and liquid water condense at 1\,bar. The constant
  graphite condensation temperature of 600\,K is because this is where
  we start our models.}
\label{SimpleModels}
\vspace*{-2mm}
\resizebox{90mm}{!}{
\begin{tabular}{c|cccc|c|cc}
          & $H$ & $O$ & $C$ & $N$ & type & $T_{\rm C}$\,[K] & $T_{\rm H2O}$\,[K]\\  
\hline
& & & &\\[-2.3ex]
simple\,1 & 0.28 & 0.48 & 0.15 & 0.09 & B & --  & 348 \\
simple\,2 & 0.45 & 0.25 & 0.15 & 0.15 & C & 600 & 358 \\
simple\,3 & 0.47 & 0.23 & 0.15 & 0.15 & C & 600 & 358 \\
simple\,4 & 0.47 & 0.21 & 0.16 & 0.16 & C & 600 & 358 \\
simple\,5 & 0.64 & 0.07 & 0.08 & 0.21 & A & --  & 337 \\
\end{tabular}}
\vspace*{-2mm}
\end{table}

The full equilibrium condensation models show that type~C atmospheres can
naturally be created by the outgassing from common rock materials such
as carbonaceous chondrites (CI) or Mid Oceanic Ridge Basalt (MORB) at
temperatures $>\!400$\,K. The two inner rocky planets of the
Trappist-1 planet system, with estimated surface temperatures of about
370\,K and 320\,K, respectively \citep{Morley2017a}, might host such
atmospheres, and will be observed with the James Webb Space Telescope, 
see details in \citet{Turbet2020}.

\smallskip
\noindent{\bf Biosignatures:\ }\ The identification of spectral
signatures of biological activity needs to proceed via two
steps. First, identify combinations of molecules which cannot
co-exist in chemical equilibrium (``non-equilibrium
markers''). Second, find biological processes that cause such
dis-equilibria, which cannot be explained by other physical
non-equilibrium processes like photo-dissociation (``biosignatures'').
The aim of this letter is to propose a robust criterion for step one.
%We have identified \ce{H2O}, \ce{CH4}, \ce{CO2} and \ce{N2} as the
%most stable set of molecules to maximse the Gibbs free energy. 
We define a non-equilibrium marker as a combination of (i) a given
molecule with (ii) one of \ce{H2O}, \ce{CH4} or \ce{CO2}, which both can
react exothermally.  Such pairs of molecules populate the opposite
corners in Fig.~\ref{fig:6mol} and have practically zero overlap in
equilibrium.  Examples of such pairs of reactants in exothermal
reactions are listed in Sect.~\ref{ChemEquil}.

\citet{KrissansenTotton2019} proposed the simultaneous detection of
\ce{CH4} and \ce{CO2} (without CO) as a biosignature, arguing that
only biological fluxes are high enough to replenish \ce{CH4} in the
upper atmospheres where it is rapidly destroyed by photochemical
processes. However, this study has shown that \ce{CH4} and \ce{CO2}
can coexist in chemical equilibrium in type~C atmosphere, which could
lead to many false positive detections.  The outgassing from warm
common rocks provides a natural mechanism to generate equilibrated
mixtures of \ce{CH4} and \ce{CO2} with only trace amounts of CO, and
at $T\!\ga\!400\,$K, type~C atmospheres may contain equally large
concentrations of \ce{H2O}, \ce{CH4} and \ce{CO2} in chemical
equilibrium.  At $T\!\approx\!350\,$K, liquid water can coexist beside
arbitrary concentrations of \ce{CH4} and a concentration of \ce{CO2}
of the order of a few percent.  It would be important to re-evaluate
photochemical effects in type~C atmospheres, where there is no free
oxygen available and, therefore, the reaction products of \ce{CH4} are
more likely to just reform \ce{CH4}.

\begin{acknowledgements}
   P.\,W.\ and Ch.\,H.\ acknowledge funding from the European Union
   H2020-MSCA-ITN-2019 under Grant Agreement no.\,860470 (CHAMELEON).
   O.\,H.\ acknowledges the PhD stipend form the University of St
   Andrews' Centre for Exoplanet Science. P.\,B.\ acknowledges support
   from the St Leonards interdisciplinary scholarship. \\*[-7mm]
\end{acknowledgements}

\bibliographystyle{aa}
\bibliography{references}

\appendix
%---------------------------------------------------------------------------
\section{Type A, B and C atmospheres}
\label{sec:types}
%---------------------------------------------------------------------------
In the following, we systematically list the principle molecular composition and
approximate abundances expected for low-temperature gases in
chemical equilibrium. These results are entirely given by the element
abundances and the stoichiometric factors of the thermodynamically
most favourable molecules, and are hence independent of pressure
and temperature.

\subsection{Type A atmospheres}
\label{typeA}
Type A atmospheres are H-rich and occur for $H>2\,O+4\,C$
(Eq.\,\ref{tri-2}). They are featured by the stability of \ce{NH3}.
Since \ce{N2} can react exothermally with \ce{H2} 
\begin{equation}
  \begin{array}{rcll}
    \ce{N2} + 3\,\ce{H2} &\longrightarrow&  2\;\ce{NH3}\hspace*{20mm}
    &:\rm -0.81\,eV  \ ,
  \end{array}
  \label{NH3reac}
\end{equation}
we have either a combination of \ce{NH3} and \ce{H2} (type~A1) or a
combination of \ce{NH3} and \ce{N2} (type~A2), depending on nitrogen abundance.

\begin{figure}
  \includegraphics[width=90mm,trim=10 4 10 11,clip]{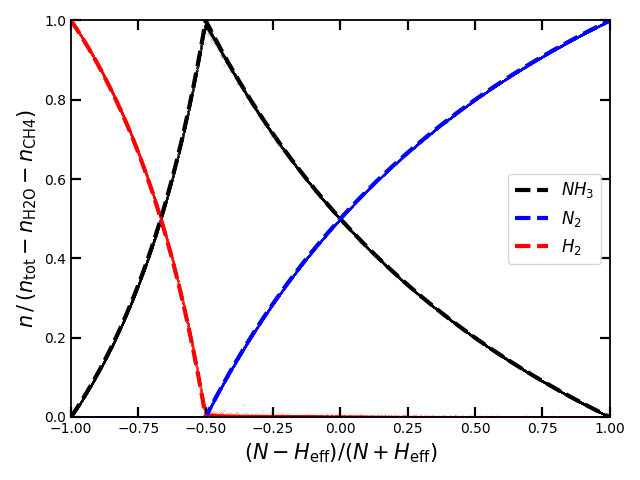}\\[-64mm]
  \hspace*{37mm}{$T\!=\!250\,$K, $p\!=\!1\,$bar}\\[59mm]
  \includegraphics[width=90mm,trim=10 4 10 11,clip]{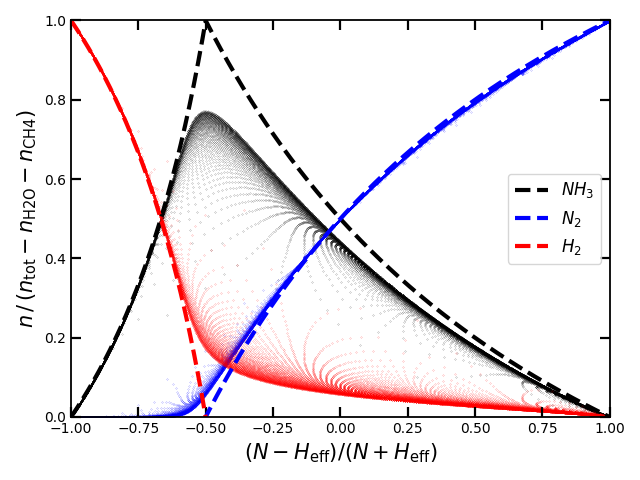}\\[-64mm]
  \hspace*{37mm}{$T\!=\!350\,$K, $p\!=\!1\,$bar}\\[55mm]
  \caption{Concentrations of \ce{NH3}, \ce{N2} and \ce{H2} in chemical
    equilibrium in type~A atmospheres. $H_{\rm eff}\!=\!H-2\,O-4\,C$ is
    the $H$ element abundance remaining after subtraction of \ce{H2O}
    and \ce{CH4}.  The dashed lines are the low-temperature
    expectations according to Eqs.~(\ref{psolA1}) and (\ref{psolA2}).
    The dots are taken from a wide range of models, systematically
    varying all $H$, $C$, $N$, $O$ element abundances at
    $p\!=\!1\,$bar, for $T\!=\!250\,$K (top plot) and for
    $T\!=\!350\,$K (bottom plot).}
  \label{Atypes}
\end{figure}

Type A1 atmospheres mainly contain \ce{H2O}, \ce{CH4}, \ce{NH3}
and \ce{H2}, and occur for $H>2\,O+4\,C$ and low nitrogen abundance $3N<H-2\,O-4\,C$.
Based on these molecules' stoichiometry, following the same procedure
as outlined in Sect.~\ref{ChemEquil}, the expected low-temperature
abundances in chemical equilibrium are
\begin{equation}
  \begin{array}{ccc}
  \displaystyle \frac{\pH2O}{\pgas} = \frac{2\,O}{H-N-2\,C}
     \hspace*{2mm} &\hspace*{-3mm},\hspace*{-3mm}&\hspace*{-5mm}
  \displaystyle \frac{\pCH4}{\pgas} = \frac{2\,C}{H-N-2\,C} \ ,\\[4mm]
  \displaystyle \frac{\pNH3}{\pgas} = \frac{2\,N}{H-N-2\,C}
     \hspace*{2mm} &\hspace*{-3mm},\hspace*{-3mm}&\hspace*{2mm}
  \displaystyle \frac{\pHH}{\pgas}  = \frac{H-2\,O-4\,C-3\,N}{H-N-2\,C}
  \ .\hspace*{-5mm}
  \label{psolA1}
  \end{array}
\end{equation}
Type A2 atmospheres mainly contain \ce{H2O}, \ce{CH4}, \ce{NH3}
and \ce{N2}, and occur for $H>2\,O+4\,C$ and high nitrogen abundance $3N>H-2\,O-4\,C$.
The expected low-temperature abundances in chemical equilibrium are
\begin{equation}
  \begin{array}{ccc}
  \displaystyle \frac{\pH2O}{\pgas} = \frac{6\,O}{H+2\,C+3\,N+4\,O}
     \hspace*{2mm} &\hspace*{-4mm},\hspace*{-3mm}&\hspace*{0mm}
  \displaystyle \frac{\pCH4}{\pgas} = \frac{6\,C}{H+2\,C+3\,N+4\,O} \ ,\\[4mm]
  \displaystyle \frac{\pNH3}{\pgas} = \frac{2\,H-8\,C-4\,O}{H+2\,C+3\,N+4\,O}
     \hspace*{2mm} &\hspace*{-4mm},\hspace*{-3mm}&\hspace*{1mm}
  \displaystyle \frac{\pN2}{\pgas}  = \frac{3\,N+4\,C+2\,O-H}{H+2\,C+3\,N+4\,O} \ .
  \label{psolA2}
  \end{array}
\end{equation}
Figure~\ref{Atypes} shows the concentrations of \ce{NH3}, \ce{H2} and
\ce{N2} in chemical equilibrium in type~A atmospheres. The left parts
of these plots, where $(N-H_{\rm eff})/(N+H_{\rm eff})<-0.5$, correspond
to type~A1, and the right parts $(N-H_{\rm eff})/(N+H_{\rm eff})>-0.5$
correspond to type~A2. We have used an effective $H$ abundance here,
$H_{\rm eff}=H-2\,O-4\,C$ in order to plot the results with regard to
the hydrogen abundance still available after \ce{H2O} and \ce{CH4}
formation. Comparison with a large sample of models, systematically
varying all $H$, $C$, $N$, $O$ abundances (see dots in
Fig.~\ref{Atypes}), shows that the low-temperature expectations
according to Eqs.~(\ref{psolA1}) and (\ref{psolA2}) require
temperatures $T\!\la\!300\,$K to be accurate, which differs from
type~B and type~C atmospheres, where the low-temperature expectations
are accurate up to $T\!\la\!600\,$K.  The reason for the earlier
occurrence of deviations of the model results from their
low-temperature expectations is that reaction (\ref{NH3reac}) is only
mildly exothermic, so the entropy terms kick in sooner when increasing
the temperature.  Thus, the low-temperature expectations in type~A
atmospheres are far less useful as compared to type~B and type~C
atmospheres, where they are more robust and hence more suitable for
classification. However, the borderline between type~C and type~A
atmospheres, and the prediction of \ce{H2O} and \ce{CH4}
concentrations remains robust even in type~A atmospheres.

Both type A1 and type A2 atmospheres can contain water clouds if
temperatures are sufficiently low, but no graphite (soot) clouds, and
only traces of \ce{CO2}, \ce{O2} and \ce{CO} molecules in chemical
equilibrium.

\subsection{Type B atmospheres}
\label{typeB}
Type B atmospheres are oxygen-rich and occur for $2\,O>H+4\,C$
(Eq.\,\ref{tri-1}). They mainly contain \ce{H2O}, \ce{CO2}, \ce{N2}
and \ce{O2}. With similar stoichiometric arguments as presented in
Sect.~\ref{ChemEquil}, the expected low-temperature abundances
in chemical equilibrium are
\begin{equation}
  \begin{array}{ccc}
  \displaystyle \frac{\pH2O}{\pgas} = \frac{2\,H}{H+2\,O+2\,N}
     \hspace*{2.5mm} &\hspace*{-3mm},\hspace*{-1mm}&
  \displaystyle \frac{\pCO2}{\pgas} = \frac{4\,C}{H+2\,O+2\,N} \ ,\\[4mm]
  \displaystyle \frac{\pN2}{\pgas} = \frac{2\,N}{H+2\,O+2\,N}
                     &\hspace*{-3mm},\hspace*{-1mm}&\hspace*{1mm}
  \displaystyle \frac{\pO2}{\pgas}  = \frac{2\,O-H-4\,C}{H+2\,O+2\,N} \ .
  \label{psolB}
  \end{array}
\end{equation}
Type B atmospheres can contain water clouds if temperatures are
sufficiently low, but no graphite (soot) clouds, and only traces of
\ce{CH4}, \ce{H2}, \ce{NH3} and \ce{CO} molecules in chemical equilibrium.

\subsection{Type C atmospheres}
\label{typeC}
Type C atmospheres have been discussed in Sect.~\ref{ChemEquil}.  They
occur in the triangle between the conditions listed by
Eqs.\,(\ref{tri-1}) to (\ref{tri-3}). The main molecules are \ce{H2O},
\ce{CO2}, \ce{CH4} and \ce{N2} with low-temperature concentrations
approximately given by Eq.\,(\ref{psol}).  Type C atmospheres can
contain water and graphite (soot) clouds if temperatures are
sufficiently low, but only traces of \ce{O2}, \ce{H2}, \ce{NH3} and \ce{CO}
molecules in chemical equilibrium.

\section{Variation of temperature, pressure, and nitrogen abundance}

\begin{figure*}
  \resizebox{182mm}{!}{
  \begin{minipage}{185mm}
  \begin{tabular}{ccc}
  \hspace*{-5mm}
  \includegraphics[height=59mm,trim=10 0 79 11,clip]{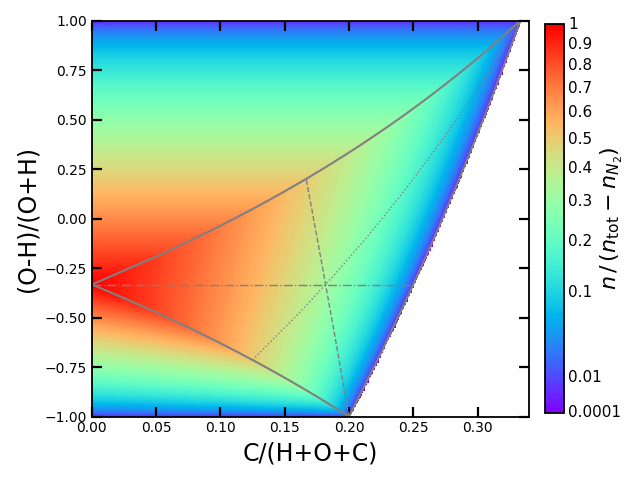} &
  \hspace*{-5mm}
  \includegraphics[height=59mm,trim=64 0 79 11,clip]{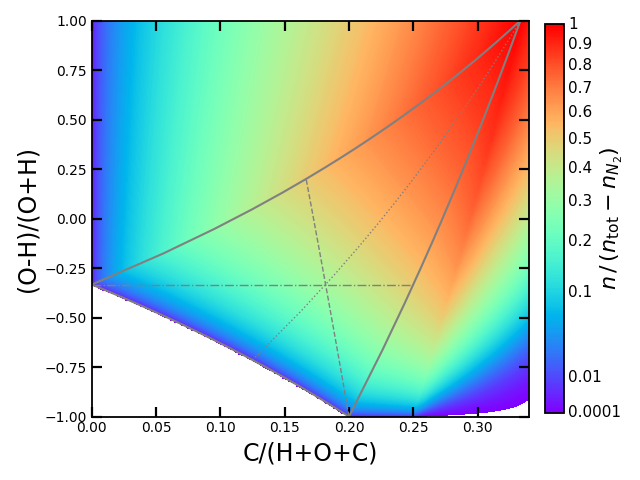} &
  \hspace*{-5mm}
  \includegraphics[height=59mm,trim=64 0  0 11,clip]{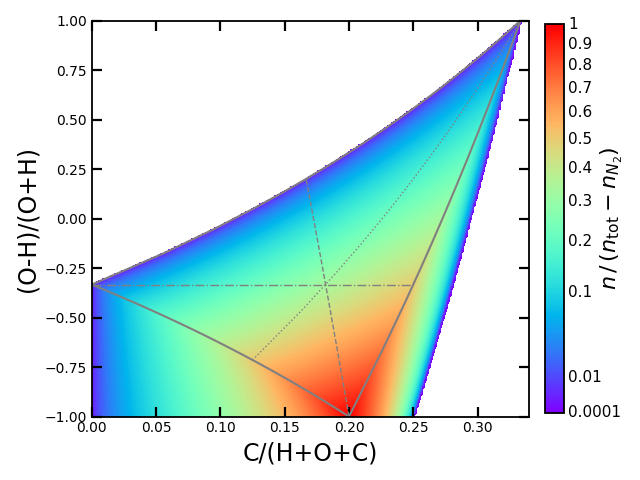} \\[-6.5mm]
  \hspace*{-5mm}
  \includegraphics[height=59mm,trim=10 0 79 11,clip]{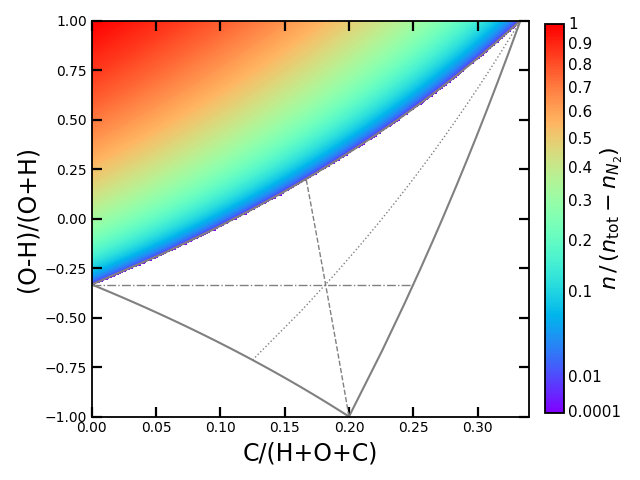} &
  \hspace*{-5mm}
  \includegraphics[height=59mm,trim=64 0 79 11,clip]{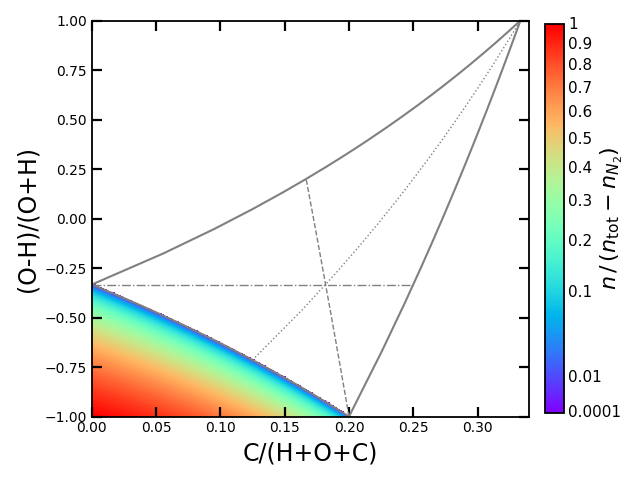} &
  \hspace*{-5mm}
  \includegraphics[height=59mm,trim=64 0  0 11,clip]{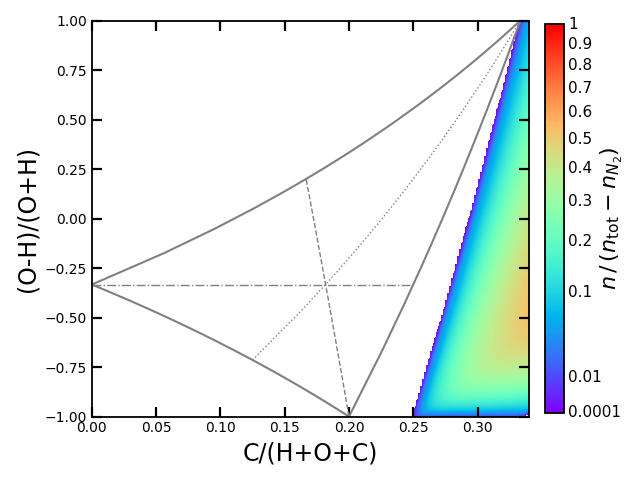} 
  \end{tabular}\\[-85mm]
  \hspace*{10mm}{\large\bff\color{white}H$_{\mathbf 2}$O}\\[-30mm]
  \hspace*{105mm}{\large\bff\color{white}CO$_{\mathbf 2}$}\\[37mm]
  \hspace*{148mm}{\large\bff\color{white}CH$_{\mathbf 4}$}\\[9mm]
  \hspace*{10mm}{\large\bff\color{white}O$_{\mathbf 2}$}\\[39mm]
  \hspace*{66mm}{\large\bff\color{white}H$_{\mathbf 2}$}\\[-13mm]
  \hspace*{168mm}{\large\bff\color{black}CO}\\[13mm]
  \end{minipage}}
  \caption{Same as Fig.~\ref{fig:6mol}, but for $T\!=\!200\,K$,
    $p\!=\!1$\,bar, and $N\!=\!0.001$.}
  \label{200K-1bar}
\end{figure*}

\begin{figure*}
  \resizebox{182mm}{!}{
  \begin{minipage}{185mm}
  \begin{tabular}{ccc}
  \hspace*{-5mm}
  \includegraphics[height=59mm,trim=10 0 79 11,clip]{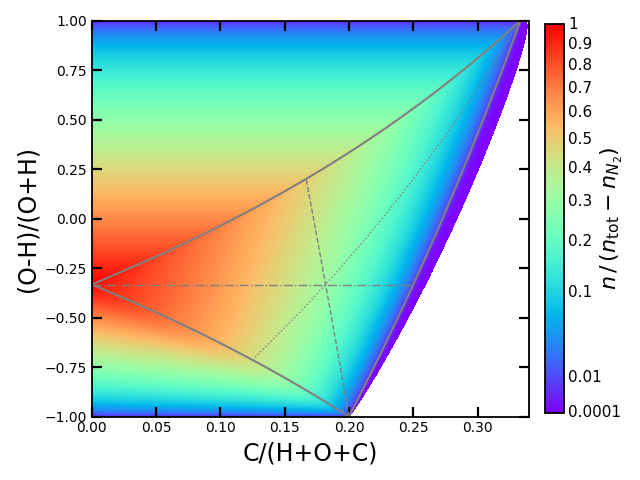} &
  \hspace*{-5mm}
  \includegraphics[height=59mm,trim=64 0 79 11,clip]{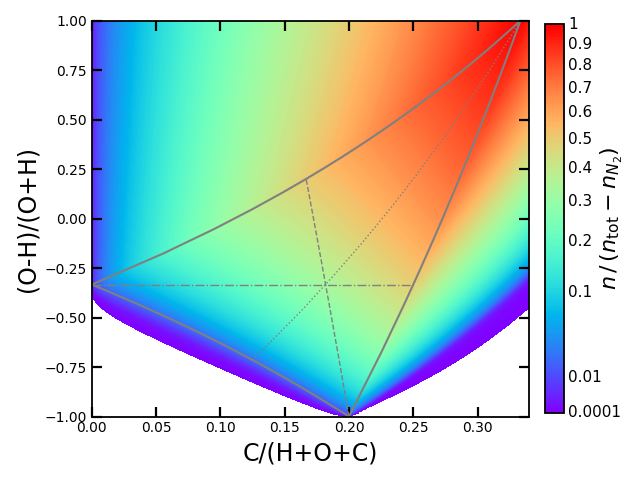} &
  \hspace*{-5mm}
  \includegraphics[height=59mm,trim=64 0  0 11,clip]{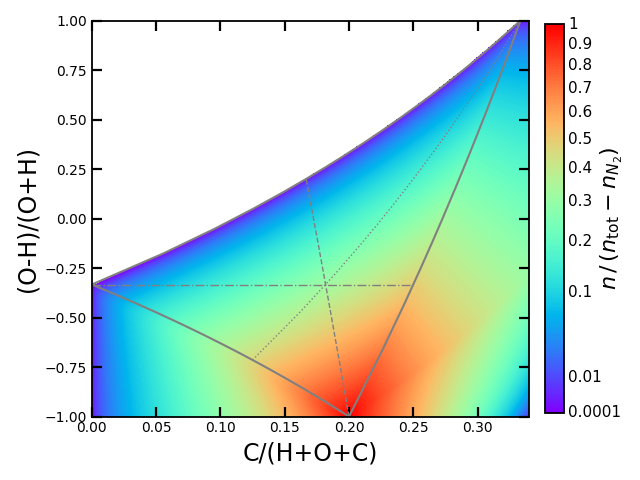} \\[-6.5mm]
  \hspace*{-5mm}
  \includegraphics[height=59mm,trim=10 0 79 11,clip]{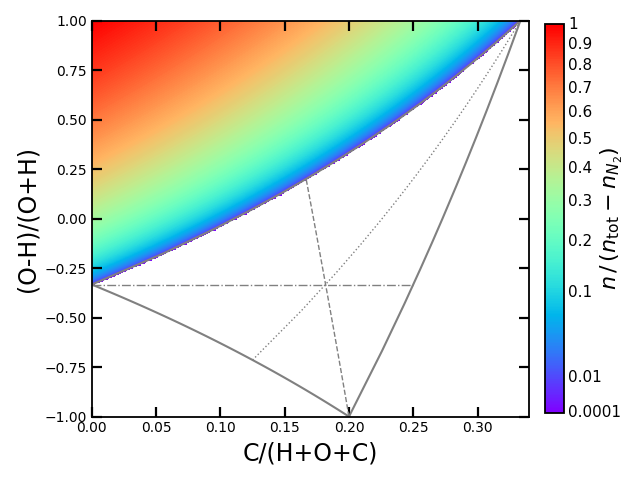} &
  \hspace*{-5mm}
  \includegraphics[height=59mm,trim=64 0 79 11,clip]{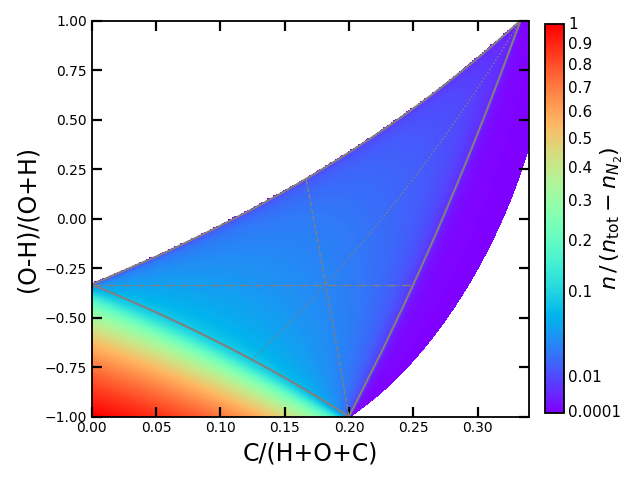} &
  \hspace*{-5mm}
  \includegraphics[height=59mm,trim=64 0  0 11,clip]{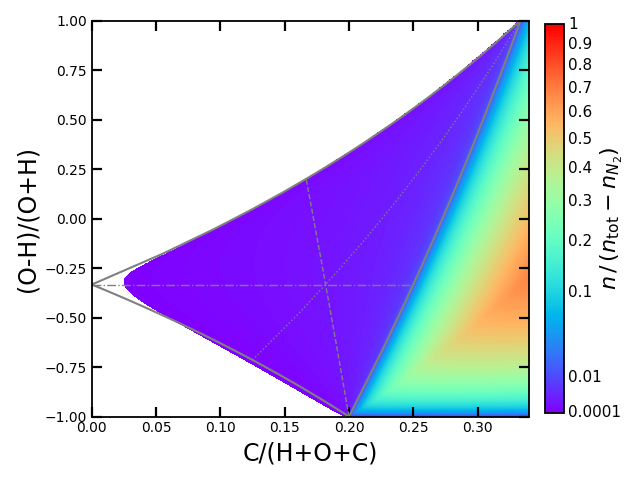} 
  \end{tabular}\\[-85mm]
  \hspace*{10mm}{\large\bff\color{white}H$_{\mathbf 2}$O}\\[-30mm]
  \hspace*{105mm}{\large\bff\color{white}CO$_{\mathbf 2}$}\\[37mm]
  \hspace*{148mm}{\large\bff\color{white}CH$_{\mathbf 4}$}\\[9mm]
  \hspace*{10mm}{\large\bff\color{white}O$_{\mathbf 2}$}\\[39mm]
  \hspace*{66mm}{\large\bff\color{white}H$_{\mathbf 2}$}\\[-13mm]
  \hspace*{168mm}{\large\bff\color{black}CO}\\[13mm]
  \end{minipage}}
  \caption{Same as Fig.~\ref{fig:6mol}, but for $T\!=\!600\,K$,
    $p\!=\!1$\,bar, and $N\!=\!0.001$.}
  \label{600K-1bar}
\end{figure*}

Figures \ref{200K-1bar} ($T\!=\!300\,$K) and \ref{600K-1bar}
($T\!=\!600\,$K) show the effects of temperature. The cold case is
very pure and the molecular concentrations are very close to
Eq.\,(\ref{psol}), with negligible concentrations of all trace
molecules in all type A, B and C atmospheres.  The warmer model shows
notable concentrations of \ce{H2} and \ce{CO} in type~C atmospheres,
but without much feedback on the main molecules, as already discussed
in Table~\ref{contaminations}.

\begin{figure*}
  \resizebox{182mm}{!}{
  \begin{minipage}{185mm}
  \begin{tabular}{ccc}
  \hspace*{-5mm}
  \includegraphics[height=59mm,trim=10 0 79 11,clip]{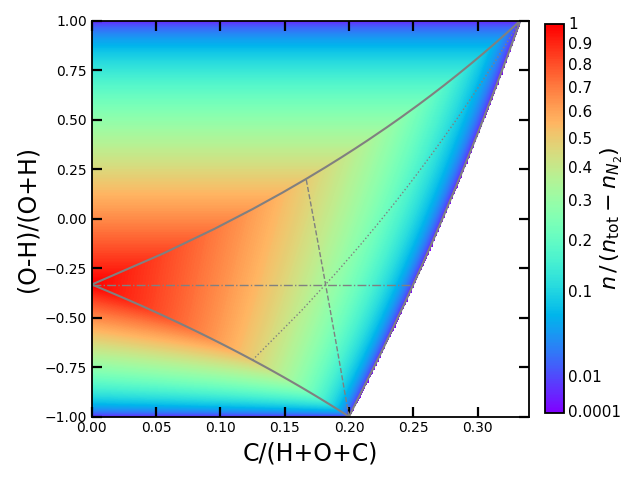} &
  \hspace*{-5mm}
  \includegraphics[height=59mm,trim=64 0 79 11,clip]{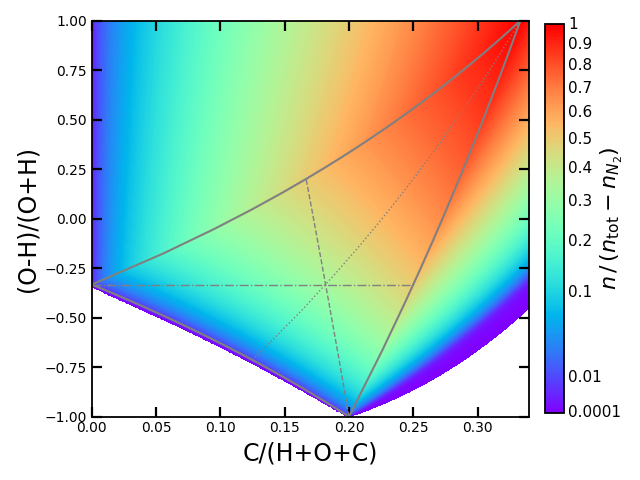} &
  \hspace*{-5mm}
  \includegraphics[height=59mm,trim=64 0  0 11,clip]{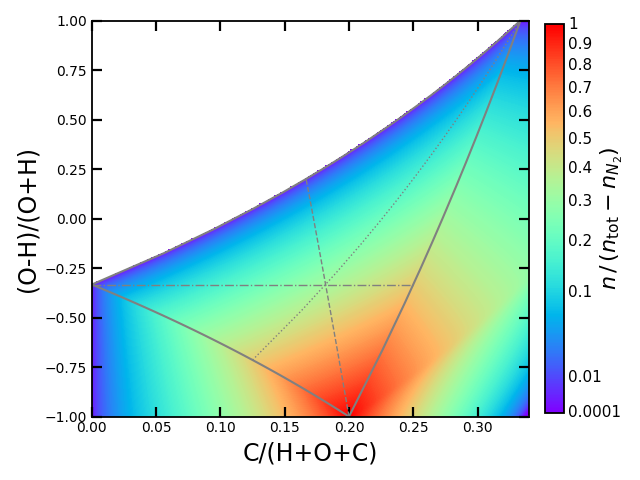} \\[-6.5mm]
  \hspace*{-5mm}
  \includegraphics[height=59mm,trim=10 0 79 11,clip]{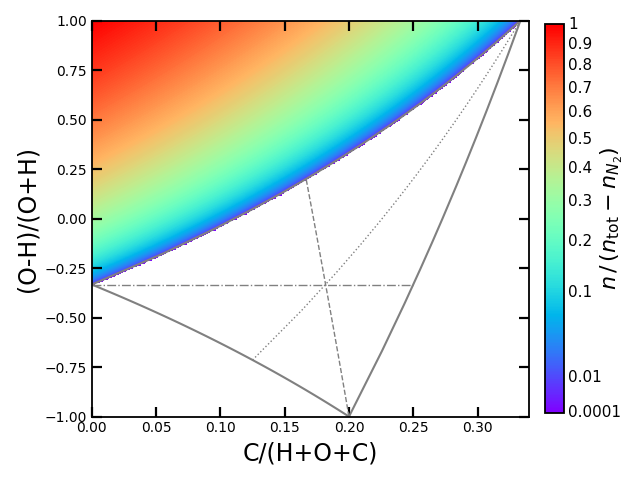} &
  \hspace*{-5mm}
  \includegraphics[height=59mm,trim=64 0 79 11,clip]{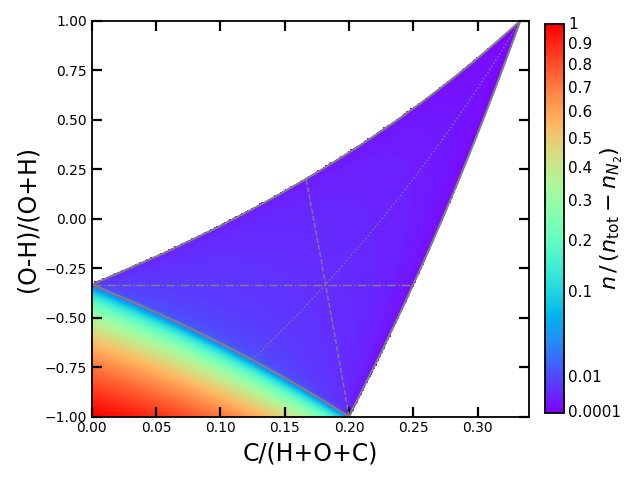} &
  \hspace*{-5mm}
  \includegraphics[height=59mm,trim=64 0  0 11,clip]{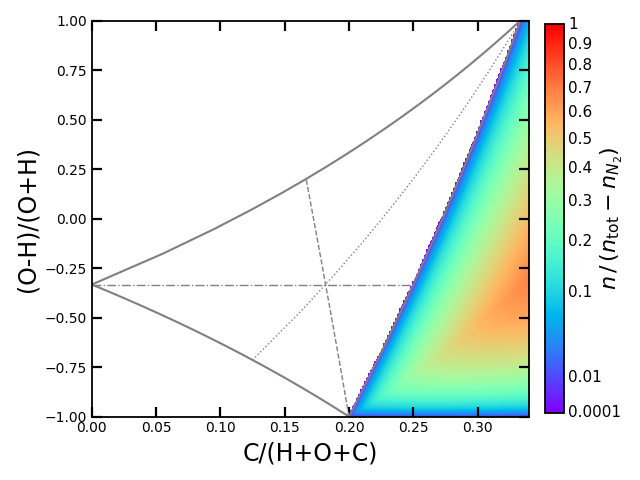} 
  \end{tabular}\\[-85mm]
  \hspace*{10mm}{\large\bff\color{white}H$_{\mathbf 2}$O}\\[-30mm]
  \hspace*{105mm}{\large\bff\color{white}CO$_{\mathbf 2}$}\\[37mm]
  \hspace*{148mm}{\large\bff\color{white}CH$_{\mathbf 4}$}\\[9mm]
  \hspace*{10mm}{\large\bff\color{white}O$_{\mathbf 2}$}\\[39mm]
  \hspace*{66mm}{\large\bff\color{white}H$_{\mathbf 2}$}\\[-13mm]
  \hspace*{168mm}{\large\bff\color{black}CO}\\[13mm]
  \end{minipage}}
  \caption{Same as Fig.~\ref{fig:6mol}, but for $T\!=\!400\,K$,
    $p\!=\!0.01$\,bar, and $N\!=\!0.001$.}
  \label{400K-0.01bar}
\end{figure*}

\begin{figure*}
  \resizebox{182mm}{!}{
  \begin{minipage}{185mm}
  \begin{tabular}{ccc}
  \hspace*{-5mm}
  \includegraphics[height=59mm,trim=10 0 79 11,clip]{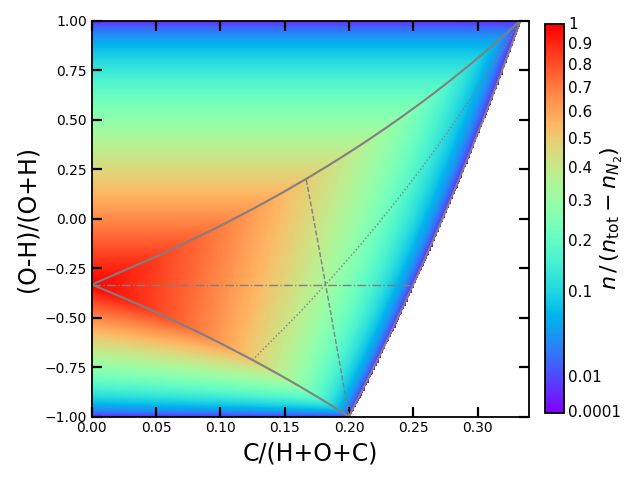} &
  \hspace*{-5mm}
  \includegraphics[height=59mm,trim=64 0 79 11,clip]{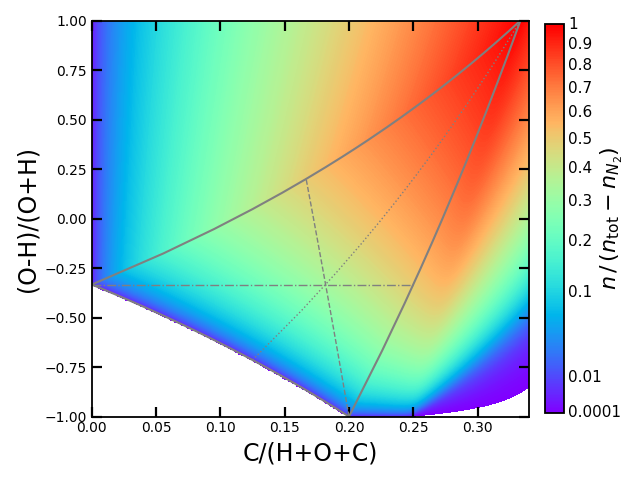} &
  \hspace*{-5mm}
  \includegraphics[height=59mm,trim=64 0  0 11,clip]{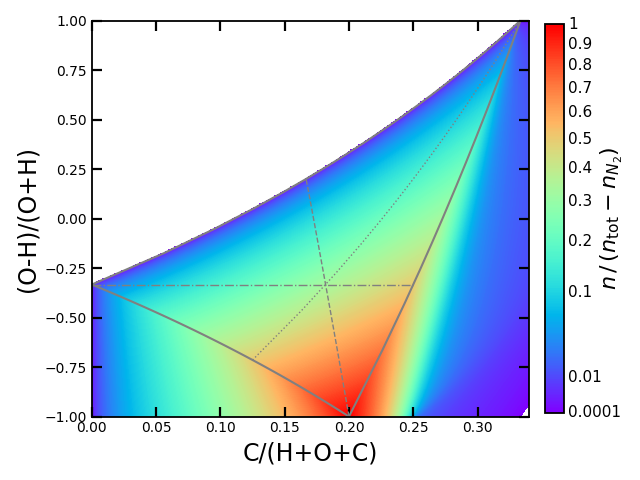} \\[-6.5mm]
  \hspace*{-5mm}
  \includegraphics[height=59mm,trim=10 0 79 11,clip]{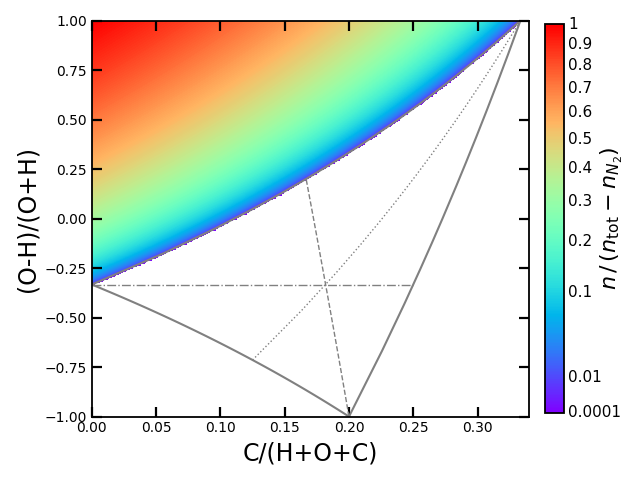} &
  \hspace*{-5mm}
  \includegraphics[height=59mm,trim=64 0 79 11,clip]{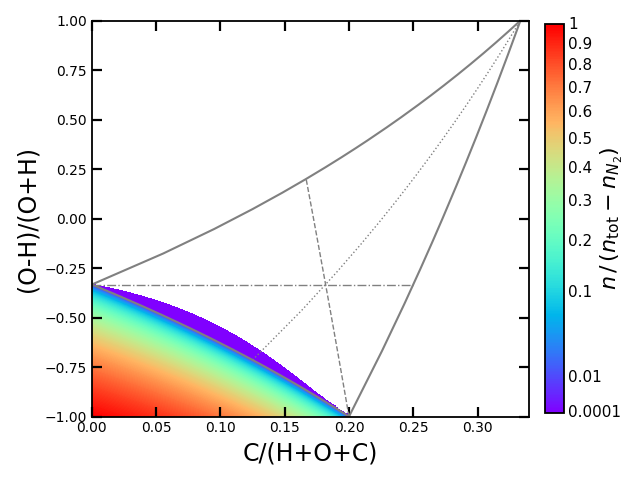} &
  \hspace*{-5mm}
  \includegraphics[height=59mm,trim=64 0  0 11,clip]{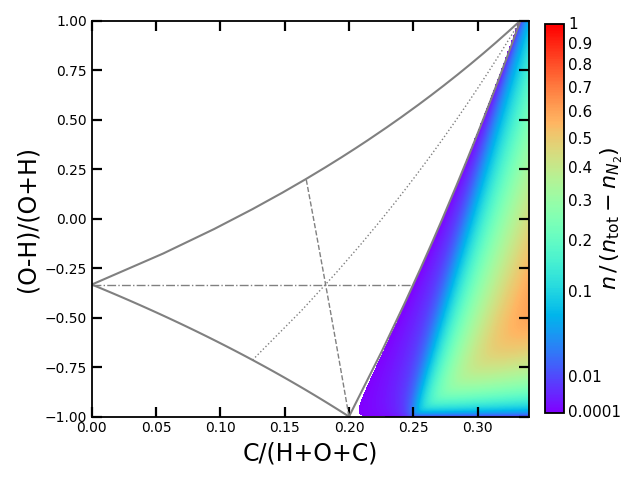} 
  \end{tabular}\\[-85mm]
  \hspace*{10mm}{\large\bff\color{white}H$_{\mathbf 2}$O}\\[-30mm]
  \hspace*{105mm}{\large\bff\color{white}CO$_{\mathbf 2}$}\\[37mm]
  \hspace*{148mm}{\large\bff\color{white}CH$_{\mathbf 4}$}\\[9mm]
  \hspace*{10mm}{\large\bff\color{white}O$_{\mathbf 2}$}\\[39mm]
  \hspace*{66mm}{\large\bff\color{white}H$_{\mathbf 2}$}\\[-13mm]
  \hspace*{168mm}{\large\bff\color{black}CO}\\[13mm]
  \end{minipage}}
  \caption{Same as Fig.~\ref{fig:6mol}, but for $T\!=\!400\,K$,
    $p\!=\!100$\,bar, and $N\!=\!0.001$.}
  \label{400K-100bar}
\end{figure*}

Figures \ref{400K-0.01bar} ($p\!=\!0.01\,$bar) and \ref{400K-100bar}
($p\!=\!100\,$bar) show that the pressure has practically no influence on
the results, except for the trace concentrations of \ce{H2} and \ce{CO} in type~C
atmospheres.

\begin{figure*}
  \resizebox{182mm}{!}{
  \begin{minipage}{185mm}
  \begin{tabular}{ccc}
  \hspace*{-5mm}
  \includegraphics[height=59mm,trim=10 0 79 11,clip]{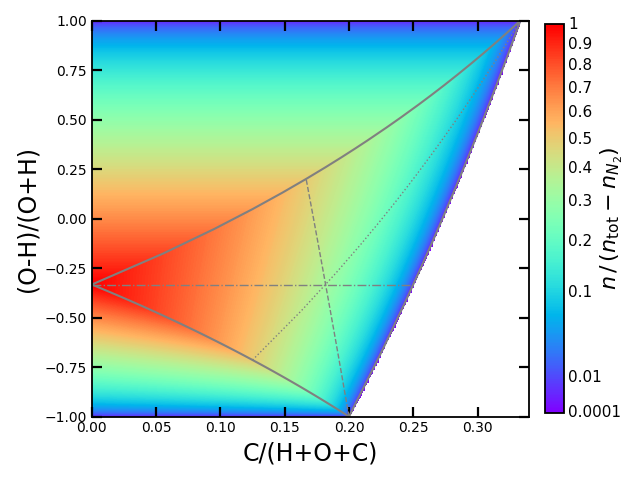} &
  \hspace*{-5mm}
  \includegraphics[height=59mm,trim=64 0 79 11,clip]{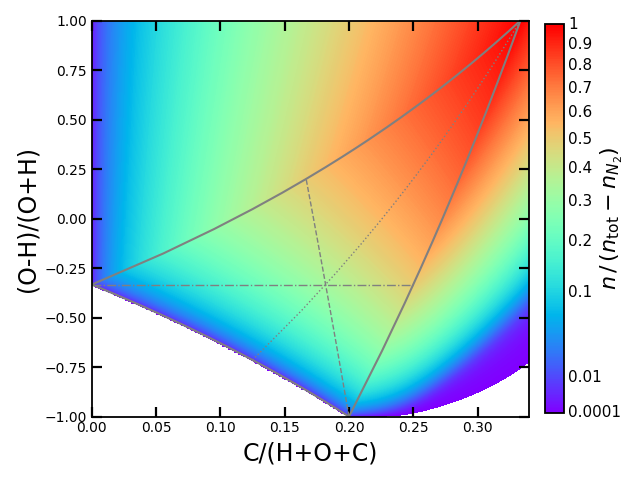} &
  \hspace*{-5mm}
  \includegraphics[height=59mm,trim=64 0  0 11,clip]{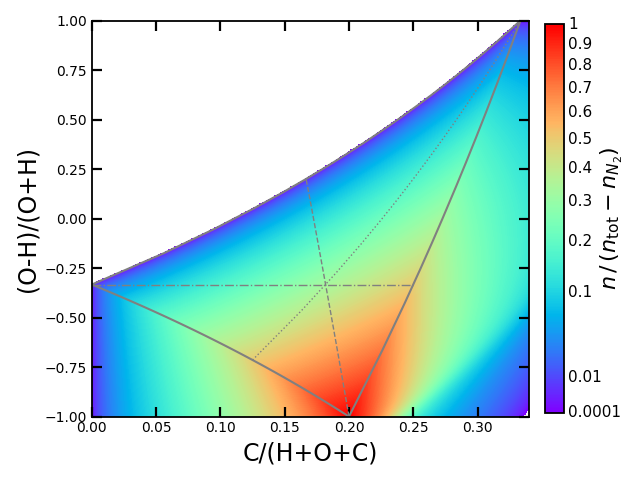} \\[-6.5mm]
  \hspace*{-5mm}
  \includegraphics[height=59mm,trim=10 0 79 11,clip]{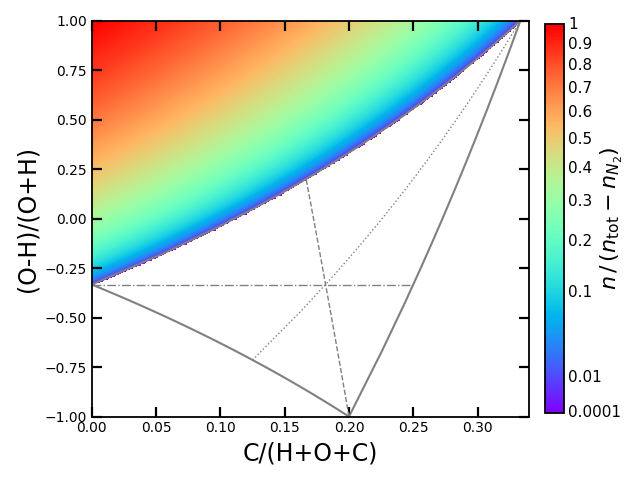} &
  \hspace*{-5mm}
  \includegraphics[height=59mm,trim=64 0 79 11,clip]{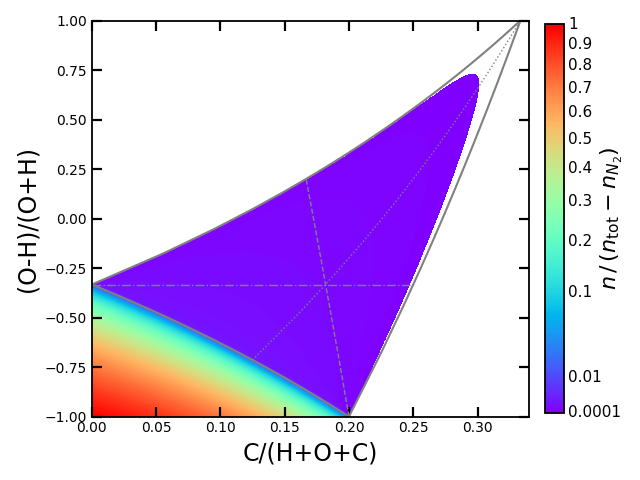} &
  \hspace*{-5mm}
  \includegraphics[height=59mm,trim=64 0  0 11,clip]{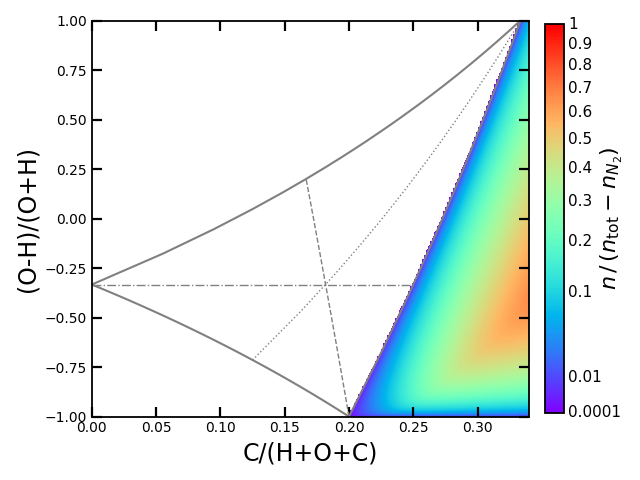} 
  \end{tabular}\\[-85mm]
  \hspace*{10mm}{\large\bff\color{white}H$_{\mathbf 2}$O}\\[-30mm]
  \hspace*{105mm}{\large\bff\color{white}CO$_{\mathbf 2}$}\\[37mm]
  \hspace*{148mm}{\large\bff\color{white}CH$_{\mathbf 4}$}\\[9mm]
  \hspace*{10mm}{\large\bff\color{white}O$_{\mathbf 2}$}\\[39mm]
  \hspace*{66mm}{\large\bff\color{white}H$_{\mathbf 2}$}\\[-13mm]
  \hspace*{168mm}{\large\bff\color{black}CO}\\[13mm]
  \end{minipage}}
  \caption{Same as Fig.~\ref{fig:6mol}, $T\!=\!400\,K$,
    $p\!=\!1$\,bar, and $N\!=\!0.001$.}
  \label{400K-1bar-N0.001}
\end{figure*}

\begin{figure*}
  \resizebox{182mm}{!}{
  \begin{minipage}{185mm}
  \begin{tabular}{ccc}
  \hspace*{-5mm}
  \includegraphics[height=59mm,trim=10 0 79 11,clip]{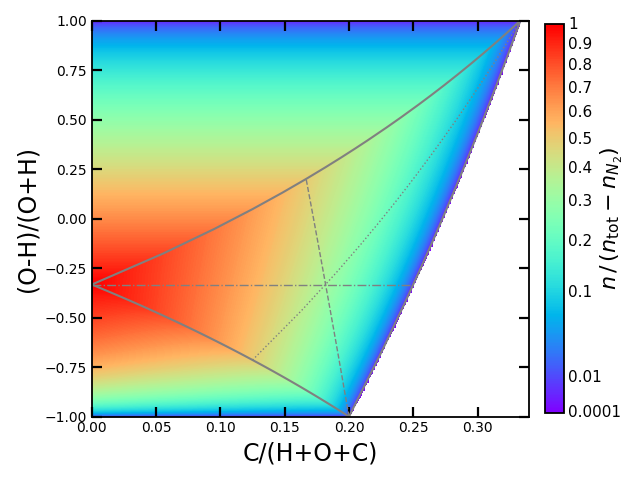} &
  \hspace*{-5mm}
  \includegraphics[height=59mm,trim=64 0 79 11,clip]{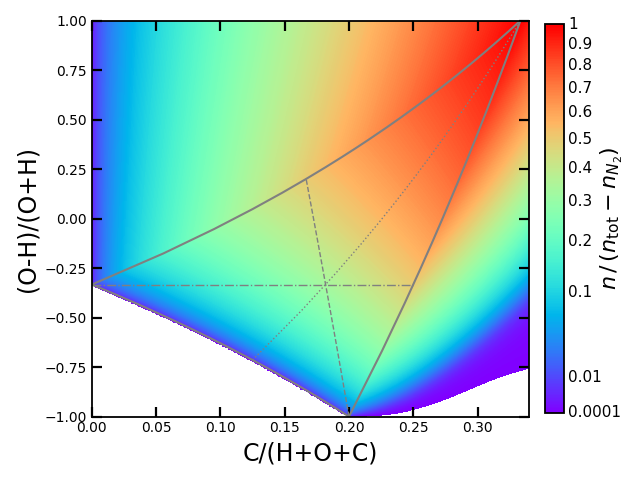} &
  \hspace*{-5mm}
  \includegraphics[height=59mm,trim=64 0  0 11,clip]{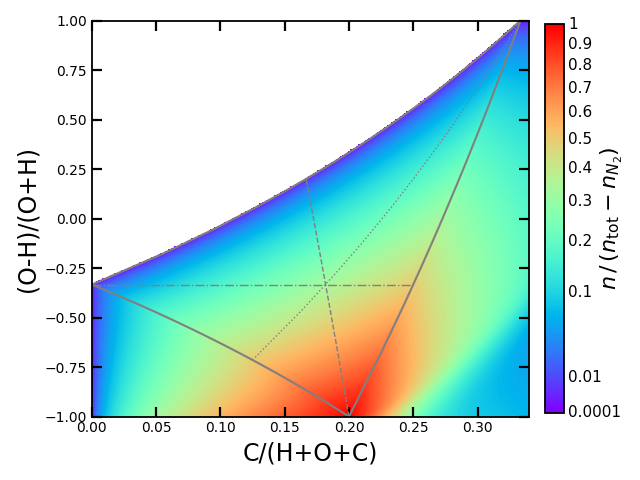} \\[-6.5mm]
  \hspace*{-5mm}
  \includegraphics[height=59mm,trim=10 0 79 11,clip]{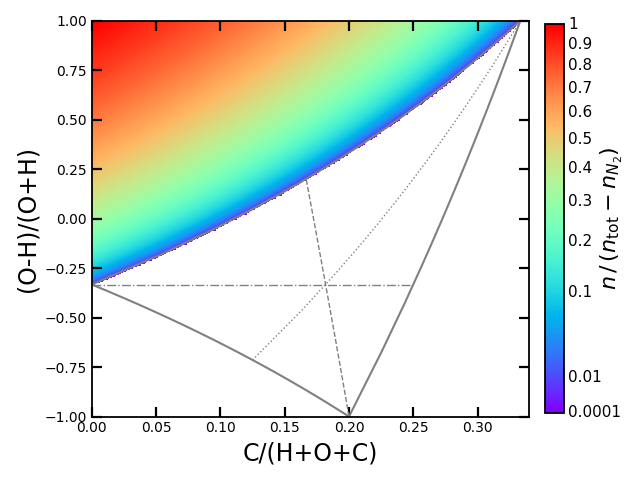} &
  \hspace*{-5mm}
  \includegraphics[height=59mm,trim=64 0 79 11,clip]{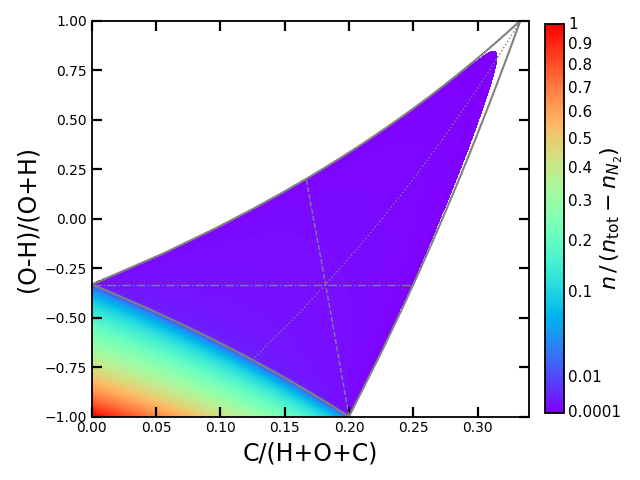} &
  \hspace*{-5mm}
  \includegraphics[height=59mm,trim=64 0  0 11,clip]{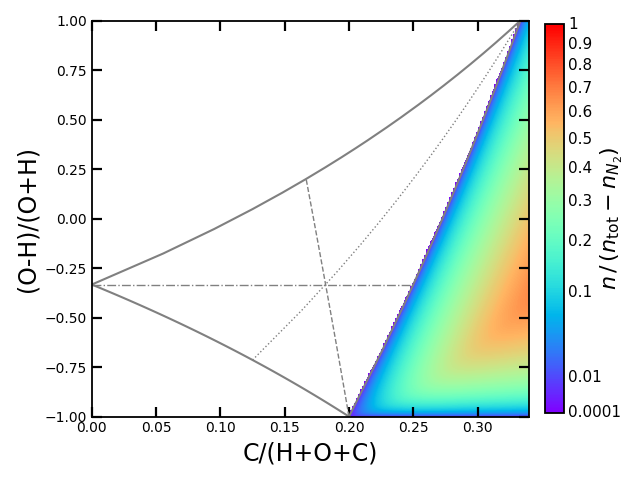} 
  \end{tabular}\\[-85mm]
  \hspace*{10mm}{\large\bff\color{white}H$_{\mathbf 2}$O}\\[-30mm]
  \hspace*{105mm}{\large\bff\color{white}CO$_{\mathbf 2}$}\\[37mm]
  \hspace*{148mm}{\large\bff\color{white}CH$_{\mathbf 4}$}\\[9mm]
  \hspace*{10mm}{\large\bff\color{white}O$_{\mathbf 2}$}\\[39mm]
  \hspace*{66mm}{\large\bff\color{white}H$_{\mathbf 2}$}\\[-13mm]
  \hspace*{168mm}{\large\bff\color{black}CO}\\[13mm]
  \end{minipage}}
  \caption{Same as Fig.~\ref{fig:6mol}, but for $T\!=\!400\,K$,
    $p\!=\!1$\,bar, and $N\!=\!0.5$.}
  \label{400K-1bar-N0.5}
\end{figure*}

Figures \ref{400K-1bar-N0.001} ($N\!=\!0.001$) and
\ref{400K-1bar-N0.5} ($N\!=\!0.5$) show the small influence of the
nitrogen abundance on the results, when we plot the concentrations
after subtraction of \ce{N2} from the total particle density. For type
B and C atmospheres, this works perfectly well, because most nitrogen
forms \ce{N2} in equilibrium, and the results for varying
$N$-abundance in type~B and C atmospheres are virtually
indistinguishable. However, for type A atmospheres, the results
actually depend on $N$, with sub-types A1 and A2 as explained in
Appendix~\ref{sec:types}, and the results would better be plotted in a
three dimensional way.  Hence, Figs.~\ref{400K-1bar-N0.001} and
\ref{400K-1bar-N0.5} only provide cuts through this 3D space, at
selected $N$-abundances, for type A atmospheres in the lower left
corner.

\end{document}